\documentclass[twoside]{article}
\usepackage{amsfonts,amssymb,amsbsy,textcomp,marvosym,amsmath,caption,threeparttable,amsthm}
\usepackage{eurosym,mathrsfs,fancyhdr,CJK,multicol,graphics,indentfirst,color,bm,upgreek,booktabs,graphicx}
\usepackage{epstopdf}
\usepackage{subfig}
\usepackage{lipsum}
\makeatletter
\usepackage{textcomp}
\usepackage{booktabs}
\usepackage{epsfig}
\usepackage{ifthen}
\usepackage{multicol}
\usepackage{bbm}
\usepackage{multirow}

\usepackage[boxed, vlined,linesnumbered]{algorithm2e}
\usepackage{color}
\usepackage{fancyhdr}
\usepackage{wrapfig}
\newcommand{\nop}[1]{}

\def\footnoterule{\kern 1mm \hrule width 10cm \kern 2mm}

\allowdisplaybreaks
\catcode`@=11
\def\title#1{\vspace{3mm}\begin{flushleft}\vglue-.1cm\Large\bf\boldmath\protect\baselineskip=18pt plus.2pt minus.1pt #1
\end{flushleft}\vspace{1mm} }
\def\author#1{\begin{flushleft}\normalsize #1\end{flushleft}\vspace*{-4pt} \vspace{3mm}}
\def\address#1#2{\begin{flushleft}\vglue-.35cm${}^{#1}$\small\it #2\vglue-.35cm\end{flushleft}\vspace{-2mm}\par}

\catcode`@=11
\def\section{\@startsection{section}{1}{\z@}%
 {-3ex \@plus -.3ex \@minus -.2ex}%
 {2.2ex \@plus.2ex}%
{\normalfont\normalsize\protect\baselineskip=14.5pt plus.2pt minus.2pt\bfseries}}
\def\subsection{\@startsection{subsection}{2}{\z@}%
 {-3ex\@plus -.2ex \@minus -.2ex}%
 {2ex \@plus.2ex}%
{\normalfont\normalsize\protect\baselineskip=12.5pt plus.2pt minus.2pt\bfseries}}
\def\subsubsection{\@startsection{subsubsection}{3}{\z@}%
 {-2.2ex\@plus -.21ex \@minus -.2ex}%
 {1.4ex \@plus.2ex}
{\normalfont\normalsize\protect\baselineskip=12pt plus.2pt minus.2pt\sl}}

\begin{document}
\begin{CJK*}{GBK}{song}
\thispagestyle{empty}
\vspace*{-13mm}
\noindent {\small W. Tian, M. Xu, G. Zhou {\it et al.} Journal of computer science and technology: Instruction for authors.
JOURNAL OF COMPUTER SCIENCE AND TECHNOLOGY \ 33(1): \thepage--\pageref{last-page}
\ September 2021. DOI 10.1007/s11390-015-0000-0}
\vspace*{2mm}

\title{Prepartition: Load Balancing Approach for Virtual Machine Reservations in a Cloud Data Center}

\author{Wenhong Tian$^{1}$, Minxian Xu$^{2,*}$, Guangyao Zhou$^{1}$, Kui Wu$^{3}$, Chengzhong Xu$^{4}$ and Rajkumar Buyya$^{1,5}$}

\address{1}{University of Electronic Science and Technology of China, Chengdu 610054, China}
\address{2}{Shenzhen Institutes of Advanced Technology, Chinese Academy of Sciences, Shenzhen 518055, China}
\address{3}{University of Victoria, Victoria, BC, V8W 3P6, Canada}
\address{4}{State Key Lab of IOTSC, University of Macau, Macau 999078, China}
\address{5}{University of Melbourne, Melbourne 3010, Australia}

\vspace{2mm}

\noindent E-mail: tian\_wenhong@uestc.edu.cn; mx.xu@siat.ac.cn; guangyao\_zhou@std.uestc.edu.cn; wkui@uvic.ca; czxu@um.edu.mo; rbuyya@unimelb.edu.au\\[-1mm]

\noindent Received July 15, 2018 [\textcolor{blue}{Month Day, Year}]; accepted October 14, 2018 [\textcolor{blue}{Month Day, Year}].\\[1mm]

\let\thefootnote\relax\footnotetext{{}\\[-4mm]\indent\ Regular Paper\\[.5mm]
\indent\ This work is supported by Key-Area Research and Development Program of Guangdong Province (NO. 2020B010164003), National Natural Science Foundation of China (NO. 62102408) and SIAT Innovation Program for Excellent Young Researchers.  \\[.5mm]
\indent\ $^*$Corresponding Author
\\[1.2mm]\indent\ $^{\footnotesize\textcircled{\tiny1}}$https://jcst.ict.ac.cn/EN/column/column107.shtml, May 2020.
\\[.5mm]\indent\ \copyright Institute of Computing Technology, Chinese Academy of Sciences 2021}

\noindent {\small\bf Abstract} \quad  {\small Load balancing is vital for the efficient and long-term operation of cloud data centers. With virtualization, post (reactive) migration of virtual machines after allocation is the traditional way for load balancing and consolidation.  However, reactive migration is not easy to obtain predefined load balance objectives and may interrupt services and bring instability. Therefore, we provide a new approach, called Prepartition, for load balancing. It partitions a VM request into a few sub-requests sequentially with start time, end time and capacity demands, and treats each sub-request as a regular VM request. In this way, it can proactively set a bound for each VM request on each physical machine and makes the scheduler get ready before VM migration to obtain the predefined load balancing goal, \color{blue}which supports the resource allocation in a fine-grained manner\color{black}. \color{black}Simulations with real-world trace and synthetic data show that Prepartition for offline (PrepartitionOff) scheduling has $10\%$-$20\%$ better performance than the existing load balancing algorithms under several metrics, including average utilization, imbalance degree, makespan and Capacity$\_$makespan. We also extend Prepartition to online load balancing. Evaluation results show that our proposed approach also outperforms existing online algorithms.}

\vspace*{3mm}

\noindent{\small\bf Keywords} \quad {\small Cloud Computing, Physical Machines, Virtual Machines, Reservation, Load Balancing, Prepartition}

\vspace*{4mm}

\end{CJK*}
\baselineskip=18pt plus.2pt minus.2pt
\parskip=0pt plus.2pt minus0.2pt
\begin{multicols}{2}

\section{Introduction}
Cloud data centers have become the foundation for modern IT services, ranging from general-purpose web services to many critical applications such as online banking and health systems. The service operator of a cloud data center is always faced with a difficult tradeoff between high performance and low operational cost \color{blue}\cite{xu2019brownout}\cite{xu2013managing}\color{black}. On the one hand, to maintain high-quality services, a data center is usually over-engineered to be capable of handling peak workload. Such up-bound configuration can bring high expenses on maintenance and energy as well as low utilization to data centers \cite{GILL2020110596}. On the other hand, to reduce cost, the data center needs to increase server utilization and shut down idle servers \cite{XU2019JSS}. The key tuning knob in making the above tradeoff is datacenter load balancing. 

Due to the importance of data center load balancing, tremendous research and development have been devoted to this domain in the past decades~\cite{zhang18}. Yet, load balancing for cloud data centers is still one of the prominent challenges that need more attention. The difficulty is compounded by several issues such as virtual machine (VM) migration, service availability, algorithm complexity, and resource utilization. The complexity in cloud data center load balancing has fostered a new industry dedicating to offer load balance services~\cite{Rahman}.

\nop{
	To deal with peak workload in traditional data centers, applications are set up on physical servers that are often over-provisioned. The up-bound configuration can bring high expenses on maintenance, energy and floor space as well as low utilization to data centers. Today’s cloud data centers are more flexible, secure and offer better support for all kinds of requests by applying virtualization. The definitions and models in this paper are general enough to be applied by different cloud providers, especially at  Infrastructure as a Service (IaaS) level. Cloud data centers are often distributed  networks in an organization,  hosting many compute nodes (such as servers), storage nodes, and network devices. Each node has multi-dimensional resources of CPUs, memory, disk storage and others.
}

Ignoring the subtle differences in detailed implementation of load balancing, let us first have a high-level view of how cloud data centers perform resource scheduling and load balancing. The process is illustrated in Fig.~\ref{fig:1}, which includes the following main steps:  
\begin{enumerate}
	\item initializing requests: user submits a VM request through a providers' web portal.
	\item matching suitable resources: based on the user's feature (such as geographic location, VM quantity and quality requirements), the scheduling center sends the VM request to an appropriate data center, in which the management program submits the request to a scheduling domain. In the scheduling domain, a scheduling algorithm is performed and resource is allocated to the request.
	\item Sending feedbacks (e.g., whether or not the request has been satisfied) to users.
	\item Scheduling tasks: determine when a VM should run on which physical machine (PM).  
	\item Optimization: the scheduling center executes optimization in the back-end and makes decisions (e.g., VM migration) for load balancing.
\end{enumerate}

\begin{center}
	\includegraphics [width=0.9\linewidth,angle=0]{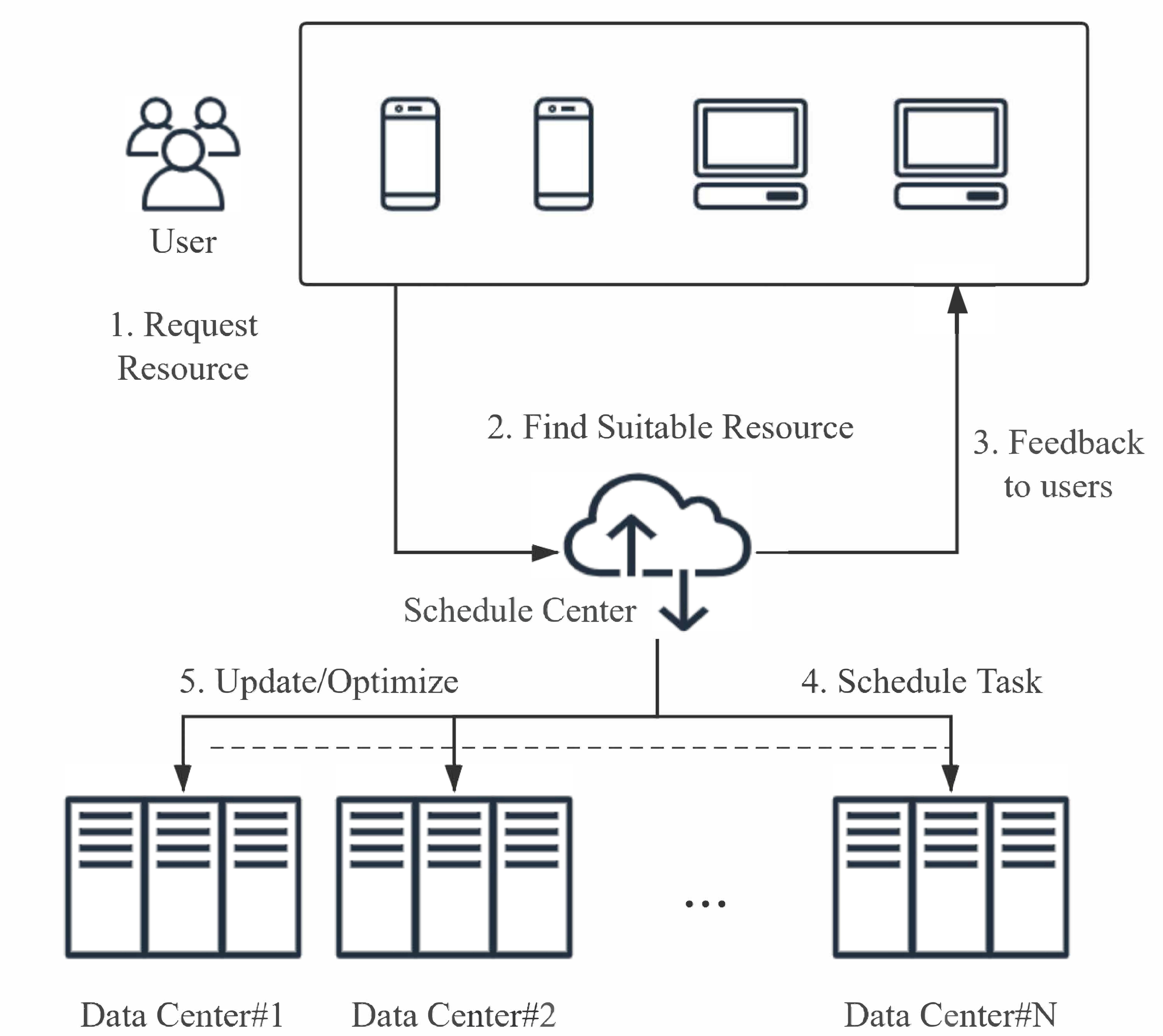}
	\parbox[c]{8.3cm}{\footnotesize{Fig.1.~}  \color{blue}A high-level view on resource scheduling/load balancing in cloud data centers.}
	\label{fig:1}
\end{center}
\setcounter{figure}{1}

\nop{
	A simplified resource scheduling process in  Cloud data centers is shown in Fig. 1, where users send requests  through Internet or intranet (Cloud). It includes the following major processes:\\
	1). User's request initialization: user initiates a request through the Internet;\\
	2). Finding suitable resources: based on the user's identity (such as geographic location, etc.) and the business characteristics (quantity and quality requirements), the scheduling center submits the request to an appropriate data center, in which the management program submits the request to a scheduling domain. In the scheduling domain, a kind of scheduling algorithm is performed and resource is allocated to the request;\\
	3). Sending feedbacks to users;\\
	4). Scheduling the tasks: executing scheduling tasks and deploying resources;\\
	5). Updating/Optimization: the scheduling center executes optimization in the back-end and prepares resources in different data centers  based on the optimization objective functions. \\
}

\color{black}	In the above process, most existing work on load balancing is reactive, i.e., performing load balancing with VM migration when unbalancing or other exceptional things happen \textit{after} VM deployment. Reactive migration of VMs is one of the practical methods for load balancing and traffic consolidation such as in VMWare. Nevertheless, it is well known that reactive VM migration is not easy to obtain predefined load balance objectives and may interrupt services and bring instability\color{blue}~\cite{zhang18}\color{black}. Our observation is that if load balancing is considered as one of the key criteria \textit{before} VM allocation, we should not only reduce the frequency of (post) VM migration (thus less service interruption), but also reach a better balanced VM allocation among different PMs.

\begin{figure*}[ht]
	\centering
	{\includegraphics[width=0.8\textwidth]{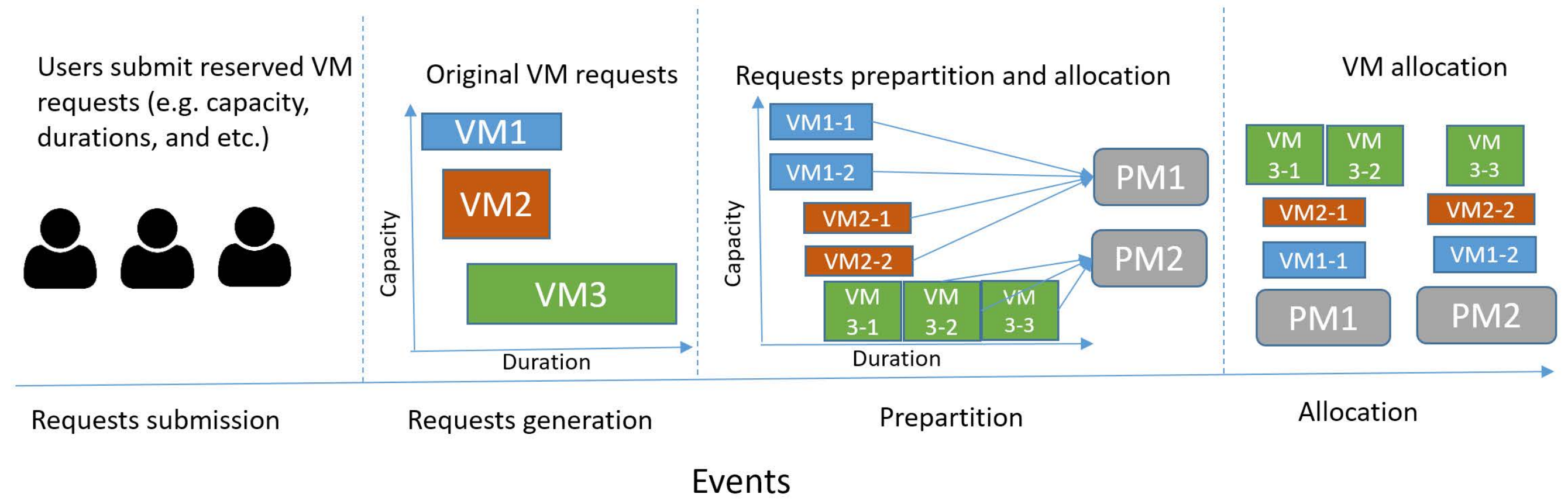}}\hfill
	
	\caption{\color{black}Illustrative example for Prepartition}
	\label{fig:events}
\end{figure*}

Motivated by the above observation, we propose a new load balancing approach called Prepartition. By combining interval scheduling and lifecycles characteristics of both VMs and PMs, Prepartition handles the problem of load balancing from a different angle. Starkly different from previous methods, it handles the VM load balance in a more proactive way. 

Fig. \ref{fig:events} shows the illustrative example based on the above observation and motivation. At the requests submission stage, the users firstly submit their reserved VM requests, including the capacity and duration information. Based on the information, then the service provider can generate the original VM request at the requests generation stage (e.g. VM1, VM2 and VM3). \color{blue}Our approach focuses on the Prepartition stage that the original VM requests can be partitioned into sub-requests and allocated to PMs before the VM migration stage, for instance, VM1 is partitioned as VM1-1 and VM1-2, and allocated to PM1. \color{black}And finally, to further optimize VM locations, VM migrations can be further applied.

\color{blue}
As the prepartition process happens before the final requests generation stage, and it does not need to execute the job, therefore, the costs are rather low compared with the overall job execution. The prepartition costs will not be the bottleneck of the system if the algorithm complexity is low. The prepartition operations can be done on the master node with powerful capability in a short time (e.g. seconds), which are much shorter compared with the execution time of jobs. 
\color{black}

The novelty of Prepartition is that it proactively sets a process-time bound (as per Capacity$\_$makespan defined in Section~\ref{sec: Problem formulation}) by \textit{pre-partitioning} each VM request and therefore helps the scheduler get ready before VM migration to achieve the predefined load balancing goal. Pre-partitioning here means that a VM request may be partitioned into a few sub-requests sequentially with start time, end time and capacity demands, where the scheduler treats each sub-request as a regular VM request and may allocate the sub-requests to different PMs\footnote{\color{black}Note that in practice we need to copy data and running state information from a VM (corresponding to a sub-request) to another VM (corresponding to the next sub-request), i.e., the operations for VM migration. But since the scheduler knows information of all sub-requests, it can prepare early so that the VM state/data transition can be finished smoothly. The implementation detail is beyond the focus of this paper.}. In this way, the scheduler can prepare in advance, without waiting for the VM migration signals as in traditional VM allocation/migration schemes. \color{blue}In addition, the resources can be allocated at the fine granularity and the migration costs can be reduced. \color{black}

To the best of our knowledge, we are the first to introduce the concept of pre-partitioning VM requests to achieve better load balancing performance in cloud data centers. This paper has the following key contributions:
\begin{itemize}
	\item Proposing a modeling approach to schedule VM reservation with sharing capacity by combining interval scheduling and lifecycles characteristics of both VMs and PMs.
	\item \color{blue}Designing novel Prepartition algorithms for both offline and online scheduling which can prepare migration in advance and set process time bound for each VM on a PM, thus the resource allocation can be made in a more fine-grained manner.\color{black}
	\item Deriving computational complexity and quality analysis for both offline and online Prepartition.
	\item Carrying out performance evaluation in terms of average utilization, imbalance degree, makespan, time costs as well as Capacity$\_$makespan  (a metric to represent loads, more details will be given in Section~\ref{sec: Problem formulation})  by simulating different algorithms with trace-driven and synthetic data.
\end{itemize}
The organization of the remaining paper is as follows: \color{black}Section~\ref{sec: Related works}  presents the related work on load balancing in cloud data centers, and Section~\ref{sec: Problem formulation} introduces problem formulation.
Section~\ref{sec: Prepartition algorithm} presents Prepartition in detail for both offline and online algorithms. Performance evaluations  are demonstrated in Section~\ref{sec: Performance evaluation}.  Finally, conclusions and future work are given in Section~\ref{sec: Conclusion and Future Directions}. \color{black}

\section{\color{black}Related work}
\label{sec: Related works}
\begin{table*}[]
	\centering
	\caption{The comparison of closely related work}
	\label{tab:relatedwork}
	\resizebox{\textwidth}{!}{%
		\begin{tabular}{|c|c|c|c|c|c|c|c|c|c|c|c|c|c|}
			\hline
			\multirow{2}{*}{\textbf{Approach}} & \multicolumn{2}{c|}{\textbf{\begin{tabular}[c]{@{}c@{}}Algorithm\\ Type\end{tabular}}} & \multicolumn{2}{c|}{\textbf{\begin{tabular}[c]{@{}c@{}}VM\\ Type\end{tabular}}} & \multicolumn{2}{c|}{\textbf{\begin{tabular}[c]{@{}c@{}}Resource \\ Type\end{tabular}}} & \textbf{\begin{tabular}[c]{@{}c@{}}Theoretical \\ Analysis\end{tabular}} & \multicolumn{6}{c|}{\textbf{Metrics}} \\ \cline{2-14} 
			& Online & Offline & Heterogeneous & \multicolumn{1}{c|}{Homogeneous} & Single & Multiple &  & Utilization & \begin{tabular}[c]{@{}c@{}}Imblance\\ Degree\end{tabular} & Makespan & \begin{tabular}[c]{@{}c@{}}Time\\ Cost\end{tabular} & \begin{tabular}[c]{@{}c@{}}Capacity\_makespan\end{tabular} & \begin{tabular}[c]{@{}c@{}}SLA \\ violations\end{tabular} \\ \hline
			Song et al. \cite{Song2015} & \checkmark &  &  & \checkmark & \checkmark &  &  & \checkmark &  &  &  &  &  \\ \hline
			Thiruvenkadam et al. \cite{Thiruvenkadam2015} & \checkmark &  & \checkmark &  &  & \checkmark &\checkmark  & \checkmark &  &  &  &  &  \\ \hline
			Wen et al. \cite{Wen2015} &  & \checkmark & \checkmark &  &  & \checkmark &  &  &  &  &  &  & \checkmark \\ \hline
			Cho et al. \cite{Cho2015} &  & \checkmark & \checkmark &  &  & \checkmark &  & \checkmark &  &  &  &  &  \\ \hline
			Tian et al. \cite{Tian2014a} &  & \checkmark & \checkmark &  &  & \checkmark & \checkmark &  & \checkmark & \checkmark &  &  &  \\ \hline
			Chhabra et al. \cite{Chhabra2019}  & \checkmark &  & \checkmark &  & \checkmark  &  &  &   \checkmark & \checkmark & & &  &  \\ \hline
			Bala et al. \cite{Bala2016} &  & \checkmark &\checkmark &  & & \checkmark &  &\checkmark & \checkmark &  & &  &  \checkmark \\ \hline
			Ebadfard et al. \cite{Ebadfard2018cpe} & & \checkmark & \checkmark & & &\checkmark &\checkmark & \checkmark & & \checkmark & & &  \\ \hline
			Ray et al. \cite{Ray2018} & & \checkmark & & \checkmark &  &\checkmark & & & \checkmark &  & & & \\ \hline
			
			\color{black}	
			Xu et al. \cite{xu2013iaware} &\checkmark &  & \checkmark &  &  &\checkmark & &\checkmark &  &  &\checkmark & & \\ \hline

			\color{black}
			Deng et al. \cite{deng2014reliability} & & \checkmark & \checkmark & &  & \checkmark & & & & &\checkmark & &\checkmark \\ \hline

			\color{black}
			Zhou et al. \cite{zhou2015carbon} &\checkmark & & & \checkmark &  \checkmark & & \checkmark & & & & & &\checkmark \\ \hline
			
			\color{black}
			Liu et al. \cite{liu2013arbitrating} &\checkmark &  & \checkmark & & \checkmark & & \checkmark & & & &\checkmark & &\checkmark \\ \hline

			Our Approach Prepartition & \checkmark & \checkmark & \checkmark &  &  & \checkmark & \checkmark & \multicolumn{1}{c|}{\checkmark} & \multicolumn{1}{c|}{\checkmark} & \multicolumn{1}{c|}{\checkmark} & \multicolumn{1}{c|}{\checkmark} & \checkmark &  \\ \hline
		\end{tabular}%
	}
\end{table*}

As introduced in several popular surveys, resource scheduling and load balancing in cloud computing have been widely studied in most works. Xu et al.~\cite{Xu2017} had a survey for the state-of-art VM placement algorithms. Ghomi et al.~\cite{Ghomi2017} recently made a comprehensive survey on load balancing algorithms in cloud computing. A taxonomic survey related to load balancing in cloud is studied by Thakur et al.~\cite{Thakur2017}. Noshy et al.~\cite{Noshy2018} reviewed the latest optimization technology dedicated to developing live VM migration. They also emphasized a further investigation, which aims to optimize the virtual machines migration process. Kumar et al. \cite{kumar2019CSUR} conducted a survey to discuss the issues and challenges associated with existing load balancing techniques. In general, approaches for VM load balancing can be categorized into two categories: online and offline. The online ones assume that only the current requests and PMs status are known, while the offline ones assume all the information is known in advance.  

\textbf{Online approach for loading balancing:} Song et al.~\cite{Song2015} proposed a VM migration method to dynamically balance VM loads for high-level application federations. Thiruvenkadam et al.~\cite{Thiruvenkadam2015} showed a hybrid genetic VMs balancing algorithm, which aims to minimize the number of migrations. Cho et al.~\cite{Cho2015} tried to maximize the balance of loads in cloud computing by combing ant colony with particle swarm optimization. 
\color{black} Xu et al. \cite{xu2013iaware} proposed iWare, which is a lightweight interference model for VM migration. The iWare can capture the relationship between VM performance interference and the important factors. 
Zhou et al. \cite{zhou2015carbon} presented a carbon-aware online approach based on Lyapunov optimization to achieve geographical load balancing. Mathematical analysis and experiments based on realistic traces have validated the effectiveness of the proposed approach. 
Liu et al. \cite{liu2013arbitrating} proposed a framework to characterize and optimize the trade-offs between power and performance in cloud platforms, which can improve operating profits while reducing energy consumption. 
\color{black}

\textbf{Offline approach for VM load balancing:} Tian et al.~\cite{Tian2014a} presented an offline algorithm on VM allocation within the reservation mode, in which the VM information is known before placement. Derived from the ant colony optimization, Wen et al.~\cite{Wen2015} proposed a distributed VM load balancing strategy with the goals of utilizing resources in a balanced manager and minimizing the number of migrations. By estimating resource usage, Chhabra et al.~\cite{Chhabra2019} developed a virtual machines placement method for loading balancing according to maximum likelihood estimation for parallel and distributed applications. Bala et al.~\cite{Bala2016} presented an approach to improving proactive load balancing by predicting multiple resource types in the cloud. In Ebadifard et al. \cite{Ebadfard2018cpe}, a task scheduling approach derived from particle swarm optimization algorithm has been proposed. In their work, tasks are independent and non-preemptive. Ray et al. \cite{Ray2018} presented a genetic-based load balancing approach to distribute VM requests uniformly among the physical machines. 
\color{black} Deng et al. \cite{deng2014reliability} introduced a server consolidation approach to achieve energy efficient server consolidation in a reliable and profitable manner. \color{black}

Different from all the above methods, 1) we investigated the reservation model that makespan and VM capacity are considered together for optimization rather than only considering the makespan or capacity separately. 2) Our approach can be applied to both online and offline scenarios rather than for a single scenario. 3) We also performed theoretical analysis for the proposed approach, and 4) evaluated more performance metrics. A qualitative comparison between our method and others is listed in Table~\ref{tab:relatedwork}. 

\section{Problem description and formulation}
\label{sec: Problem formulation}

\color{black}VMs reservation is considered that users submit their VM requests by specifying required capacity and duration. The VM allocations  are  modeled as a fixed processing time problem with modified interval scheduling (MISP).\color{black} Details on traditional interval scheduling problems with fixed processing time  were  introduced in \cite{IEEEhowto:Kleinberg2} and its references. In the following, a general formulation of  the  modified interval-scheduling problem is introduced and evaluated against some known algorithms. 
\color{black}The key symbols used throughout this work are summarized in Table \ref{tab:symbols}. \color{black}

\begin{table*}[]
	\color{black}
	\centering
	\caption{\color{black}Key notations in our models}
	\label{tab:symbols}
	\resizebox{0.8\textwidth}{!}{%
		\begin{tabular}{|c|l|c|l|}
			\hline
			\textbf{Notations} & \multicolumn{1}{c|}{\textbf{Definitions}} & \textbf{Notations} & \multicolumn{1}{c|}{\textbf{Definitions}} \\ \hline
			$T$ & The whole observation time period & $sl_0$ & The length of each time slot \\ \hline
			$n$ & The maximum number of requests & $s_i$ & The start time of request $i$ \\ \hline
			$f_i$ & The finishing time of request $i$ & $A(i)$ & The set VMs requests scheduled to PM $i$ \\ \hline
			$d_j$ & The capacity demand of VM $j$ & $CM_j^r$ & The capacity\_makespan of VM request $j$\\ \hline
			$CM_i$ & The capacity\_makespan of PM $i$ & $PCPU_i$ & The CPU capacity of PM $i$ \\ \hline
			$PMem_i$ & The memory capacity of PM $i$ & $PSto_i$ & The storage capacity of PM $i$ \\ \hline
			$VCPU_j$ & The CPU demand of VM $j$ & $VMem_j$ & The memory demand of VM $j$ \\ \hline
			$VSto_j$ & The storage demand of VM $j$ & $T_j^{start}$ & The start time of VM request $j$ \\ \hline
			$T_j^{end}$ & The finishing time of VM request $j$ & $T_r$ & The time span between time slot $t_{r-1}$ and $t_{r}$ \\ \hline
			$CM^p$ & The maximum capacity\_makespan of all PMs & $IMD$ & The imbalance degree \\ \hline
			$k$ & The partition value & $P_0$ & The lower bound of the optimal solution OPT \\ \hline
			$B_d$ & The dynamic balance value based on capacity\_makespan & $L$ & The amount of VM requests that alreadly arrived \\ \hline
			$m$ & The number of PMs in use & $I$ & A set of VM requests \\ \hline
			$CM_b$ & The predefined capacity\_makespan threshold for partition & $f$ & Constant value to avoid too frequent partitions \\ \hline
		\end{tabular}%
	}
\end{table*}

\subsection{Assumptions}
The key assumptions are:

1) The time is given in a discrete fashion; all data is given deterministically. The whole time period [0, $T$] is partitioned into equal-length $(sl_0)$, and the total number of slots is then $t$=$T/sl_0$. The start time $s_i$ and end time
$f_i$ are integer numbers of one slot. Then the interval of demand can be expressed in slot fashion with (start time, end time). For instance, if $sl_0$=10
minutes, an interval (5, 12) represents that it starts at the 5th time slot and finishes at the 12th time slot. The duration of this demand is
(12-5)$\times$10=70 minutes.

\color{blue}	2) For all VM requests generated by users, they have the start time and end time to represent their life-cycles, and the capacity to show the required amount of resources. \color{black}

2) The capacity of a single physical machine is normalized to be 1 and the required capacity of a VM can be 1/8, 1/4, or 1/2 or other portions of the total capacity of a PM. This is consistent with applications in Amazon EC2
\cite{IEEEhowto:Amazon} and \cite{Knauth2012}. 

\subsection{Key Definitions}
A few key definitions are given here:

\textit{\textbf{Definition 1}. Traditional interval scheduling problem (TISP) with fixed processing time}:
A batch of demands $\{$1, 2,$\ldots$, $n\}$ where the $i$-th demand refers to an interval of time starting at $s_i$ and finishing at $f_i$ \color{black}($\forall i, s_i < f_i$)\color{black}, each one requires
a capacity of 100\%, i.e. utilizing the full capacity of a server during that interval.

\textit{\textbf{Definition 2}. Interval scheduling with capacity sharing (ISWCS)}: 
Difference from TISP, ISWCS can share the capacities among demands if the sum of all demands scheduled on the
single server at any time is still not fully utilized.

\textit{\textbf{Definition 3}. Compatible sharing  intervals for ISWCS, for short, CSI-ISWCS}:
A batch of intervals with  requested capacities below the whole capacity of a PM during the intervals can be compatibly scheduled on a PM.
Compared against ISWCS, the requests in CSI-ISWCS can be modelled as the ones with lifecycles, which can be represented as sharing the subset of intervals.

In  the  existing literature, \textit{makespan}, i.e., the maximum total load (processing time) on any machine, is applied to measure load balancing. 

In this paper, we aim to solve the problem based on the ISWCS manner and apply a new metric Capacity$\_$makespan. 

\color{black}	\textit{\textbf{Definition 4.} The Capacity$\_$makespan of a PM $i$}: In the schedule of VM requests to PMs, denote $A(i)$ as the set of VM requests scheduled to
$PM_i$. With this scheduling, $PM_i$ will have load as the sum of  the  product of each requested capacity and its duration, called
Capacity$\_$makespan, abbreviated as $CM$, \color{black}as follows:
\begin{equation}
	CM_i=\sum_{j\in A(i)}d_j t_j
\end{equation}

in which $d_j$ is the capacity demand (some portion of total capacity) of $VM_j$ from a PM where the capacity can be CPU or Memory or storage in this paper, it can be simplified as a capacity based on Assumption 3), and $t_j$ for the span of demand $j$, being the length of processing time of demand $j$. \\
Similarly, the Capacity$\_$makespan of a given VM request is simply the  product of the requested capacity and its duration. 
\subsection{Optimization Objective}
Then, the objective of load balancing is to  minimize  the maximum load (Capacity$\_$makespan) on all PMs as noted in Eq. (2). 
Considering $m$ PMs are in the data center, we can form the problem as:
\begin{align}
	& \min (\mathop{\max}\limits_{1\le i\le m} (CM_i))\\
	&\text{subject to}~ \forall ~slot~s, \sum_{j\in A(i)} d_{j} \leq 1 
\end{align}

\noindent in which $d_j$ is the capacity demand of VM $j$ and the whole capacity of a PM $i$ is normalized to 1. \color{black}The condition (3) shows the sharing capacity
constraint that in any time interval, the shared resources should not use up all the provisioned resources (100\%). \color{black}

From this form, we see that lifecycle and capacity sharing are  key differences from traditional metrics like makespan which focuses on
process time. Traditionally Longest Process Time first (LPT) \cite{Graha} is widely adopted to load balance offline multi-processor scheduling.
Reactive migration of VMs is another way to compensate after allocation.
\subsection{Metrics for ISWCS load balancing}
A few key metrics for ISWCS load balancing are given in the following. Other metrics are the same as given in \cite{IEEEhowto:Tian4}.\\
1) PM resources: \\$PM_i(i,PCPU_i$, $PMem_i$, $PSto_i)$, $PCPU_i$, $PMem_i$, $PSto_i$ is respectively the CPU, memory, storage capacity of that a PM can
offer.\\
2) VM resources: \\$VM_j(j, VCPU_j$, $VMem_j$, $VSto_j$, $T_j^{start}$, $T_j^{end})$, $VCPU_j$, $VMem_j$, $VSto_j$ is respectively the CPU, memory, storage demand of
$VM_j$, $T_j^{start},T_j^{end}$ is respectively the start time and end time.\\
3) Discrete time: Considering a time span  be partitioned into equal length of slots. The $s$ slots can be considered as $[(t_0, t_1),(t_1,
t_2),\ldots,(t_{s-1},t_{s})]$, each time slot $T_r$ represents the time span $(t_{r-1}, t_{r})$.\\
4) Average CPU utilization of $PM_i$ during slot 0 and $T_s$ is defined as:
\begin{equation}
	PCPU_i^U=\frac{\sum_{r=0}^{s} (PCPU_i^{T_r}\times T_r)}{\sum_{r=0}^{s} T_r}
\end{equation}
where $PCPU_i^{T_r}$ is the average CPU utilization monitored and computed in slot $T_r$ which may be a few minutes long, and $PCPU_i^{T_r}$ can be obtained by monitoring CPU utilization in slot $T_r$. Average memory utilization ($PMem_i^U$) and storage utilization ($PSto_i^U$) of PMs can
be calculated similarly. Similarly, the average CPU (memory and storage) utilization of a VM can be calculated.\\
5) Makespan: represents the whole length of the scheduled VM reservations, i. e., the difference between the start time of the first request \footnote{in this paper, we interchange demands and requests, both of them are referred to VM requests}  and the finishing time of
the last request.\\
6) The maximum Capacity$\_$makespan ($CM^{p}$) of all PMs: is calculated as:
\begin{align}
	CM^p=& \max_i{(CM_i)}
\end{align}
where we can apply CPU, memory and storage utilization too. \\
7) Imbalance degree (IMD): The variance is a metric of how far a set of values are spread out from each other in statistics. IMD is the normalized variance (regarding its average) of CPU, memory and storage utilization for all PMs and it measures load imbalance effect and is defined as:
\begin{align}
	\frac{\sum_{i=0}^{m} (\frac{(Avg_i-CPU_u)^2}{3}+\frac{(Avg_i-Mem_u)^2}{3}+\frac{(Avg_i-Sto_u)^2}{3})}{m}
\end{align}
where 
\begin{align}
	Avg_i=\frac{PCPU_i^U+PMem_i^U+PSto_i^U}{3} 
\end{align}
and $CPU_u$, $Mem_u$,$Sto_u$ is respectively the average utilization of CPU, memory and storage in a Cloud data center during consideration and can be computed using utilization of all PMs in a Cloud data center.  


\textbf{ Theorem 1.}  Minimizing the makespan in the offline scheduling problem is NP-hard. \\
The proof was provided in our previous work \cite{IEEEhowto:Tian4} and is omitted here. Our model in this paper differs from the previous one in several perspectives: 1). we model that the multiple VM requests are allowed to be executed on the same host simultaneously rather than a single VM request; 2) Our objective is  minimizing the Capacity\_makespan rather than the longest processing time; 3) we extend our model to be suitable for both offline and online scenarios.


Combining the properties of both fixed process time intervals and capacity sharing, we present new offline and online algorithms in the following section.

\section{Prepartition Algorithms}
\label{sec: Prepartition algorithm}

In the following, we introduce one algorithm for the offline scenario and two algorithms for the online scenario, which can handle both the offline and online requests, and achieve good performance in load balancing. 

\subsection{PrepartitionOff Algorithm}

\color{blue}First, we introduce the PrepartitionOff algorithm that aims to partition the VM requests under the situation that the information of all VM requests is known in advance. In this way, the processed order of VM requests and the prepartition operations can be managed by the algorithm. 

\color{black}Considering a set of VM reservations, there are $m$ PMs in a data center and denote $OPT$  as the optimal solution with regard to minimizing the
Capacity$\_$makespan. Firstly define
\begin{equation}
	P_0= \frac{1}{m}\sum_{j=1}^{J} CM_j^r \leq OPT
\end{equation}
where $J$ denotes the total number of allocated VMs, and $P_0$ denotes the lower bound for the $OPT$.

Algorithm \ref{PrepartitionOff} gives the pseudocodes of PrepartitionOff algorithm which measures the ideal load balancing among $m$ PMs. The algorithm firstly calculates the balancing
value by formula (9), sets a partition value ($k$) and computes the length of each partition, i.e. $\lceil P_0/k\rceil$, representing the maximum CM that a VM can continuously
be allocated on a PM (line 1). 
For every demand, PrepartitionOff divides it into multiple $\lceil P_0/k\rceil$ subintervals when its CM is equal to or larger than
$P_0$, and each subinterval is treated as a new request (lines 2-4).
Then the algorithm sorts the newly generated requests in decreasing order based on $CM$ for further scheduling (line 5).
After sorting of requests, the algorithm will pick up the VM with the earliest start time, and allocates the VM to the PM with the lowest average Capacity$\_$makespan and enough available resources (lines 6-8), thus achieving the load balancing objective. The Capacity$\_$makespan  of PM will also be updated accordingly (line 9).
Finally, the algorithm calculates the Capacity$\_$makespan of each PM when all requests are assigned and finds total partition numbers (line 10). In practice, the scheduler records all
possible subintervals and their hosting PMs so that migrations of VMs can be prepared beforehand to alleviate overheads.
\color{black}

\begin{algorithm*}[!tb]
	\SetArgSty{textnormal}
	\caption{The pseudo codes of PrepartitionOff algorithm}\label{PrepartitionOff}
	\KwIn{$m$:total number of PMs; $n$: total number of VM requests, requests are indicated by their (demanded VM IDs, start times, finishing times, demanded capacity), \color{black}$CM_j^r$ is the Capacity$\_$makespan of request $j$\color{black}, $CM_i$ for the Capacity$\_$makespan of PM $i$;}
	\KwOut{Assign PM IDs to all requests and their partitions}
	Initialization:  computing the bound value $P_0$ and partition value $k$ (e.g. 1, 2, ...); 
	
	\ForAll{$i$ from $1$ to $m$}
	{\If{ $CM_i \geq P_0$ }
		{divide it by $\lceil{P_0/k}\rceil$ subintervals equally and treat each subinterval as a new request}
	}
	All intervals are sorted in decreasing order of $CM$, break ties arbitrarily;
	
	
	\ForAll{$j$ from $1$ to $n$} {\nllabel{PartitionOff Outer Loop Begin Line}
		Pick up the VM with the earliest start time in the VM queue for execution;\
		
		Allocate $j$ to the PM with the smallest load and enough capacity;\
		
		Update load $(CM)$ of the PM;}
	
	
	
	Calculate CM of every PM and total number of partitions
\end{algorithm*}

\textbf{ Theorem 2.} Applying priority queue data structure, the PrepartitionOff algorithm has a computational complexity of $O(nlogm)$, where $n$ is the number of VM requests after pre-partition and $m$ is the total number of PMs used. \\
\noindent Proof: The priority queue is adopted so that each PM has a priority value (average Capacity$\_$makespan), and each time the algorithm chooses an item from it, the algorithm selects the one with the highest priority.
It costs $O(n)$ time to sort $n$ elements, and $O(logn)$ steps for insertion and the extraction of minima in a priority queue \cite{IEEEhowto:Kleinberg2}. Then, by adopting a priority queue, the algorithm picks a PM with the lowest average Capacity$\_$makespan in $O(logm)$ time. In total, the time complexity of the PrepartitionOff algorithm is $O(nlogm)$ for $n$ demands. 

\textbf{ Theorem 3.}  PrepartitionOff algorithm has approximation ratio of  $(1+\epsilon)$ regarding the capacity\_makespan where $\epsilon$=$\frac{1}{k}$ and $k$ is the partition value (a preset constant).\\

Proof: It can be seen that every demand has bounded Capacity$\_$makespan by Preparition applying the lower bound $P_0$. Every
request has start time $s_i$, end time $f_i$ and process time $p_i$=$f_i$-$s_i$. Think the last job (later than all other jobs) to complete and assume the start time of this job is $T_0$.  \color{black}We also assume that all other servers are allocated with VM requests and denote the maximum Capacity$\_$makespan as $CM_m$, this means $CM_m\leq$OPT. \color{black}Since, for all requests $i \in J$ , we have
$CM_i\leq\epsilon$OPT (by the setting of PrepartitionOff algorithm in formula  (9)),  this job finishes with load ($CM_m$+$\epsilon$OPT). Therefore, the schedule with
Capacity\_makespan ($CM_m$+$\epsilon$OPT) $\leq$ (1+$\epsilon$)OPT, this ends the proof.   

\subsection{PrepartitionOn1 Algorithm}
\color{blue}Apart from the offline scenario, the online scenario is also quite common in a realistic environment. For online VM allocations, scheduling decisions must be made without complete information about the entire job instances because jobs arrive one by one. We firstly extend the offline Prepartition algorithm to the online scenario as PrepartitionOn1, which can only have the information of VM requests when the requests come into the system.\color{black}

Given $m$ PMs and $L$ VMs (including the one that just came) in a data center. Firstly define

\begin{equation}
	B_d=min({\max_{1\le j\le L} {(CM_j^r)/2}, {\sum_{j=1}^{L} (CM_j^r)}/m})
\end{equation}
\color{black}$B_d$ is called dynamic balance value, which is one-half of the max Capacity$\_$makespan of all current PMs or the ideal load balance value of all current PMs in the system, where $L$ is the number of VMs requests already arrived. Notice that the reason to set $B_d$ as one half of the max Capacity$\_$makespan of all current
PMs is to avoid that a large number of partitions may cause extra management costs.

Algorithm 2 shows the pseudocodes of the PrepartitionOn1 algorithm. Since in an online algorithm, the requests come one by one, the system can only capture the
information of arrived requests. 
The algorithm firstly predefines the prepartition value $k$ and the total partition number $P$ as 0 (line 1).  
When a new request comes into the system, the algorithm picks up the VM with the earliest start time in the queue for scheduling and computes dynamic balance value ($B_d$) by equation (10) (lines 3-4). 
After $B_d$  is computed, if Capacity$\_$maskespan of VM request is too large (larger than $\lceil(B_d/k)\rceil$), then the
initial request is partitioned into several requests (segments) based on the partition value $k$. In these partitioned requests,  if some requests are still with large Capacity$\_$maskespan, they would be put back into the queue waiting to be executed, and follow the same partition and allocation process (lines 5-6). The VM requests with small  Capacity$\_$maskespan after partition would be executed
when their start time begins, which will be assigned to the PM that has the lowest value of  Capacity$\_$makespan (lines 7-9). 
Once all demands are allocated, PrepartitionOn1
calculates the Capacity$\_$makespan value of all the PMs and outputs all the partition values for $n$ demands (line 10). Since the number of partitions and segments of each VM request are known at the moment of allocation, the system can prepare VM migration in advance so that the processing time and instability of migration can be reduced.
\color{black}

\begin{algorithm*}[!tb]
	\SetArgSty{textnormal}
	\caption{PrepartitionOn1 Algorithm}\label{PrepartitionOn1}
	\KwIn{$m$: total number of PMs; $n$: total number of VM requests, requests are indicated by their (demanded VM IDs, start times, finishing times, demanded capacity), \color{black}$CM_j^r$ is the Capacity$\_$makespan of request $j$\color{black}, $CM_i$ for the Capacity$\_$makespan of PM $i$;}
	\KwOut{Assign PM IDs to all requests and their partitions}
	Set the partition value $k$, total partition number $P$=0;
	
	
	\For {each arrived job $j$} {\nllabel{PrepartitonOn1 Outer Loop Begin Line}
		\color{black}			Pick up the VM with the start time equals to system time in the VM queue to schedule\color{black};
		Compute $CM_j^r$ of VM $j$, and $B_d$ using Eq.~(10)\;
		\If{$CM_j^r>\lceil(B_d/k)\rceil$} {partition $VM_j$ into multiple $\lceil(B_d/k)\rceil$ equal subintervals, treat each subinterval as a new demand and add them
			into VM queue,~$P=P+\lceil \frac{CM_j^r}{B_d/k}\rceil$, Update load $(CM)$ of the PM\;
			\nllabel{End If}
		}
		
		\Else{Allocate $j$ to PM with the smallest load and enough capacity\;
			Update load $(CM)$ of the PM\;}
	}\nllabel{PrepartitonOn1 Outer Loop End Line}
	Output total number of partitions $P$\
\end{algorithm*}

\color{black}	To analyze algorithm performance based on theoretical analysis, we conduct \textit{competitive ratio} analysis that represents the performance ratio between an online algorithm and an optimal offline algorithm. An online algorithm is competitive if its competitive ratio is bounded. 

\textbf{Theorem 4.} PrepartitionOn1 has a competitive ratio of $(1+\frac{1}{k}-\frac{1}{mk})$ with regarding to the Capacity$\_$makespan. \\
\color{black}
Proof: Without loss of generality, we label PMs in order of non-decreasing final loads (CM) in PrepartitionOn1. Denote $OPT$ and $PrepartitionOn1(I)$ as the optimal load balance value of corresponding offline scheduling and load balance value of PrepartitionOn1 for a given set of jobs $I$, respectively. Then the load of $PM_m$
defines the Capacity$\_$makespan. The first $(m$-1) PMs each process a subset of the jobs and then experience an (possibly none) idle period. All PMs together
finish a total Capacity$\_$makespan $\sum_{i=1}^{n} CM_i$ during their busy periods. Consider the allocation of the last job $j$ to PM$_m$. By the scheduling rule of PrepartitionOn1, PM$_m$ had the lowest load at the time of allocation.  Hence, any idle period on the first ($m$-1) PMs cannot be bigger than the
Capacity$\_$makespan of the last job allocated on PM$_m$ and hence cannot exceed the maximum Capacity$\_$makespan divided by $k$ (partition value), i.e.,
$\frac{\max_{1\leq i\leq n}CM_i}{k}$. Based on Equation (10), then we have \\
\begin{equation}
	m\times PrepartitionOn1(I)\leq \sum_{i=1}^{n} (CM_i) +(m-1)\frac{\max (CM_i)}{k}
\end{equation}
which is equivalent to
\begin{equation}
	PrepartitionOn1(I)\leq \sum_{i=1}^{n}(\frac{CM_i}{m})+(m-1)\frac{\max(CM_i)}{mk}
\end{equation}
which is
\begin{equation}
	PrepartitionOn1(I)\leq (OPT+(\frac{1}{k}-\frac{1}{mk})OPT)
\end{equation}
Note that $\frac{\sum_{i=1}^{n} CM_i}{m}$ is the lower bound on $OPT(I)$ because the optimum Capacity$\_$makespan cannot be smaller than the average
Capacity$\_$makespan on all PMs. And $OPT(I) \geq \max_{1\leq
	i\leq n}CM_i$ since the largest job must be processed on a PM. We therefore have $PrepartitionOn1(I)\leq (1+\frac{1}{k}-\frac{1}{mk})OPT$.

\textbf{Theorem 5.} By using the priority queue, the computational complexity of PrepartitionOn1 is $O(nlogm)$, where $n$ is the number of VM requests after the pre-partition operations and $m$ is the total number of used PMs. \\
\\
\noindent Proof: It is similar to the proof for Theorem 2, therefore, we omit it here. \\
\subsection{PrepartitionOn2 Algorithm}
\color{black}
\color{blue}Observing that the PrepartitionOn1 may bring too many partitions in some cases, we present the PrepartitionOn2 algorithm by introducing a parameter to control the number of partitions in a more flexible manner.
 \color{black}The differences from
PrepartitionOn1 are the followings:\\
1) To avoid a large number of partitions, we bring a constant value $f$~(for instance 0.125, 0.25 and etc.)~ for measuring load balancing;\\
2) Setting a CM bound for each PM, for instance, each PM has a CM=1$\times$24 in each day within 24 hours, but we consider a PM can at most run with 100$\%$ CPU utilization in 16
hours, i.e., we set a CM bound for each PM for each day as $CM_B$=16.
If overloading happens to a PM according to predefined thresholds $(1+f)$ and $CM_B$, then a new request should be partitioned into multiple $x$ (the number of active PMs) subintervals
equally and the scheduler allocates each subinterval to every PM.

\color{black}
The pseudocodes of the PrepartitionOn2 algorithm are shown in Algorithm 3. The algorithm firstly initializes the predefined Capacity\_makespan bound of PMs and the constant value $f$ as introduced above (line 1). For the arrived VMs, the algorithm picks up the VM with the earliest start time for execution, and calculates the Capacity\_makespan of both VMs and PMs (lines 2-4). The picked VM will be supposed to be allocated to the PM with the smallest Capacity\_makespan value, and the Capacity\_makespan of PM as well as the PM with the smallest Capacity\_makespan are calculated with that supposition (lines 5-7). If the increased Capacity\_makespan of PM is too large (line 8), the VM will be partitioned into the number of active PMs, and the partitioned VMs are allocated to PMs one by one (line 9). Otherwise, the VM can be allocated directed to the PM with the smallest loads (lines 10-11). Finally, the scheduling results and number of partitions can be obtained (line 12).   
\color{black}

\color{black}

\textbf{Theorem 6.} PrepartitionOn2 has a computational complexity of $O(nlogm)$ by applying  a  priority queue, where $n$ is the number of VM requests after
the pre-partition operations and $m$ is the total number of used PMs. \\
\noindent Proof: It is also similar to the proof for Theorem 2, we therefore omit it. 

\color{black}\textbf{Theorem 7.} The competitive ratio of PrepartitionOn2 is at most $(1+f)$ and each PM has CM at most $CM_B$ with regard to the Capacity$\_$makespan. \\
Proof: According to Algorithm \ref{PrepartitionOn2}, whenever a PM has CM larger than $CM_B$ or the competitive ratio of the algorithm is larger than $(1+f)$; the allocating VMs will be pre-partitioned into multiple sub-instances and allocated. Therefore the competitive ratio of PrepartitionOn2 is at most (1+$f$). This completes the proof.\color{black}

\begin{algorithm*}[!tb]
	\SetArgSty{textnormal}
	\caption{PrepartitionOn2 Algorithm}\label{PrepartitionOn2}
	\KwIn{$m$: total number of PMs; $n$: total number of VM requests, requests are indicated by their (demanded VM IDs, start times, finishing times, demanded capacity), $CM_j^r$ is the Capacity$\_$makespan of request $j$, $CM_i$ for the Capacity$\_$makespan of PM $i$;}
	\KwOut{Assign PM IDs to all requests and their partitions}
	Initialization: set the CM bound $CM_B$ for each PM, a constant value $f(\leq0.125)$, total partition number $P$=0;
	
	
	\For {each arrived job $j$} {\nllabel{PrepartitonOn2 Outer Loop Begin Line}
		Pick up the VM with the earliest start time in the VM queue to schedule\;
		Compute $CM_j^r$ of VM $j$, and $CM$ of each PM\;
		Choose the minimum value $CM$ of PM named $CM_{oldmin}$\;
		Suppose to allocate the VM $j$ to the PM which has $CM_{oldmin}$ and compute its' new value of $CM$ named $CM_{oldmin^+}$\;
		Get the new minimum value of $CM$ of PM named $CM_{newmin}$\;
		\If{(${CM_{oldmin^+}/CM_{newmin})}>{(1+f)}$ or $CM_{oldmin^+} > CM_B$}
		{partition $VM$ $j$ into multiple $x$ (the number of PM turned on) subintervals equally, allocate each subinterval to every PM, ~$P=P+x$ \;
			\nllabel{End If}
			\Else{Allocate $j$ to PM with the lowest load and available capacity\;
		}}
	}\nllabel{PrepartitonOn2 Outer Loop End Line}
	Output total number of partitions $P$\
\end{algorithm*}

\section{Performance Evaluation}
\label{sec: Performance evaluation}

\begin{table*}
	\scriptsize
	\caption{8 types of VMs derived from Amazon EC2}
	\begin{center}
		\begin{tabular}{|c|c|c||c|}
			\hline Compute Capacity (Units)& Memory (GB) & Storage (GB)& VM Type
			\\\hline
			\hline 1 & 1.875 & 211.25 & 1-1(1) \\
			\hline 4  & 7.5  & 845 & 1-2(2) \\
			\hline 8  & 15 & 1690 &1-3(3) \\
			\hline 6.5 & 17.1  & 422.5 &2-1(4)\\
			\hline 13  & 34.2 & 845 &2-2(5)\\
			\hline 26  & 68.4 & 1690 &2-3(6)\\
			\hline 5  & 1.875 & 422.5 &3-1(7) \\
			\hline 20  & 7 & 1690 &3-2(8)\\
			\hline
		\end{tabular} \\
	\end{center}
	\label{tab:ECSspecification}
\end{table*}

\begin{table*}
	\caption{3 types of suggested PMs specification
	}
	\begin{center}
		\scriptsize
		\begin{tabular}{|c|c|c||c|}
			\hline PM Type& Compute Capacity (Units) & Memory (GB) & Storage (GB)
			\\\hline
			\hline 1 & 16  & 30 & 3380 \\
			\hline  2 & 52  & 136.8 & 3380 \\
			\hline  3 & 40  & 14 & 3380\\
			\hline
		\end{tabular} \\
	\end{center}
	\label{tab:PMspecification}
\end{table*}
\color{black}	Notice that there are three types of PMs in Table \ref{tab:PMspecification} and 8 types VMs in Table \ref{tab:ECSspecification}, where each type of VM occupies 1/16 or 1/8 or 1/4 or 1/2 of the whole capacity of the corresponding PM considering all three dimension resources of CPU, memory, and storage, therefore the three-dimension resources become one dimension in this case. In the future, we will extend to other cases.\\ 
In the following, the simulation results of Prepartition algorithms and a few existing algorithms are provided. To conduct simulation, a Java simulator called CloudSched (refer to Tian et al.~\cite{IEEEhowto:Tian2}) is used.
\color{black}

\subsection{Offline Algorithm Performance Evaluation}

All simulations ran on a computer configured with an Intel i5 processor at 2.5GHz and 4GB memory. All VM requests are generated by following Normal
distribution.
In offline algorithms, Round-Robin (RR) algorithm, Longest Process Time (LPT) algorithm and Post Migration Algorithm (PMG) are implemented and compared. 
\begin{figure*}[ht]
	\centering
	\subfloat[]{{\includegraphics[width=0.24\textwidth]{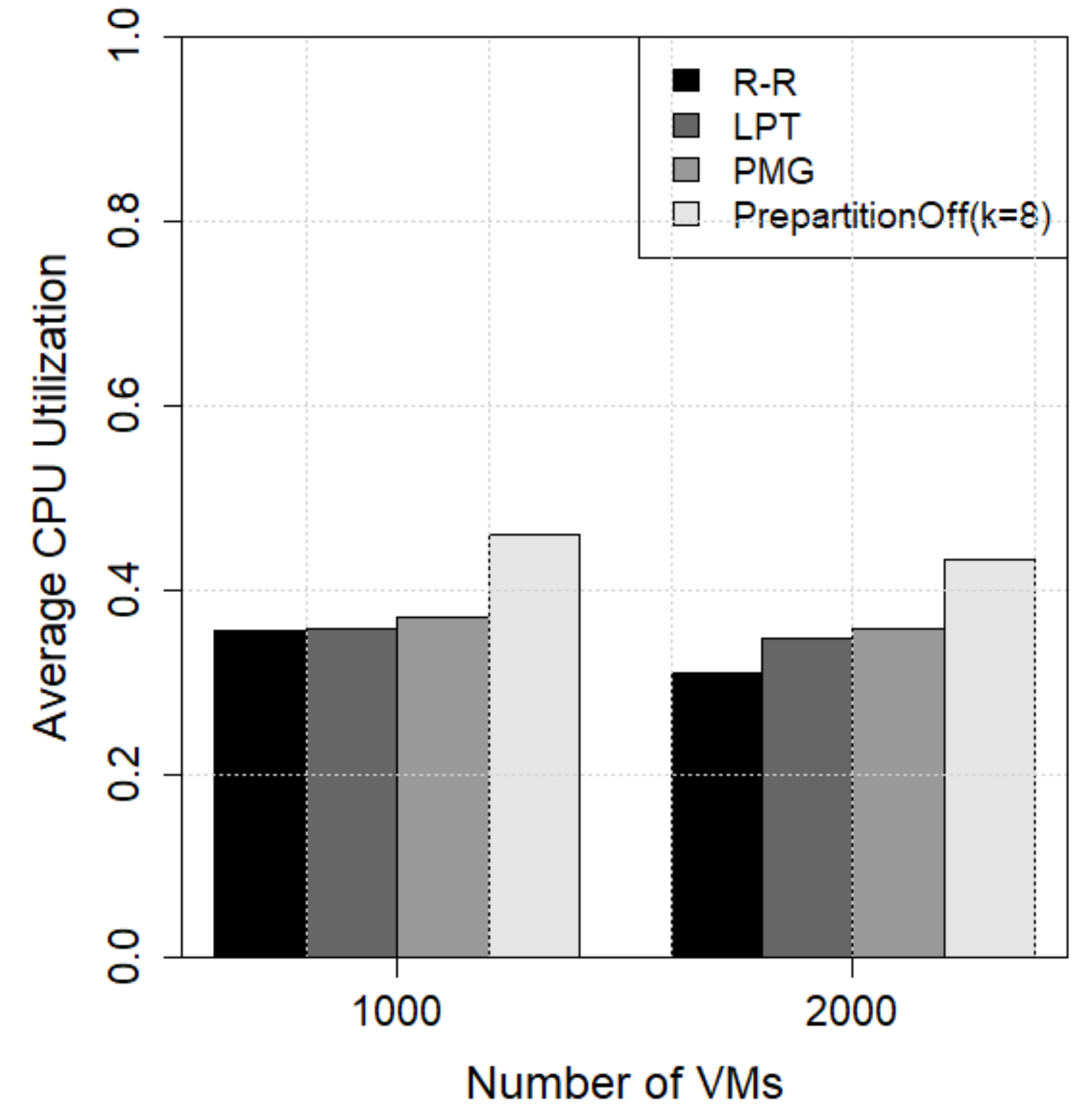}}}\hfill
	\subfloat[]{{\includegraphics[width=0.24\textwidth]{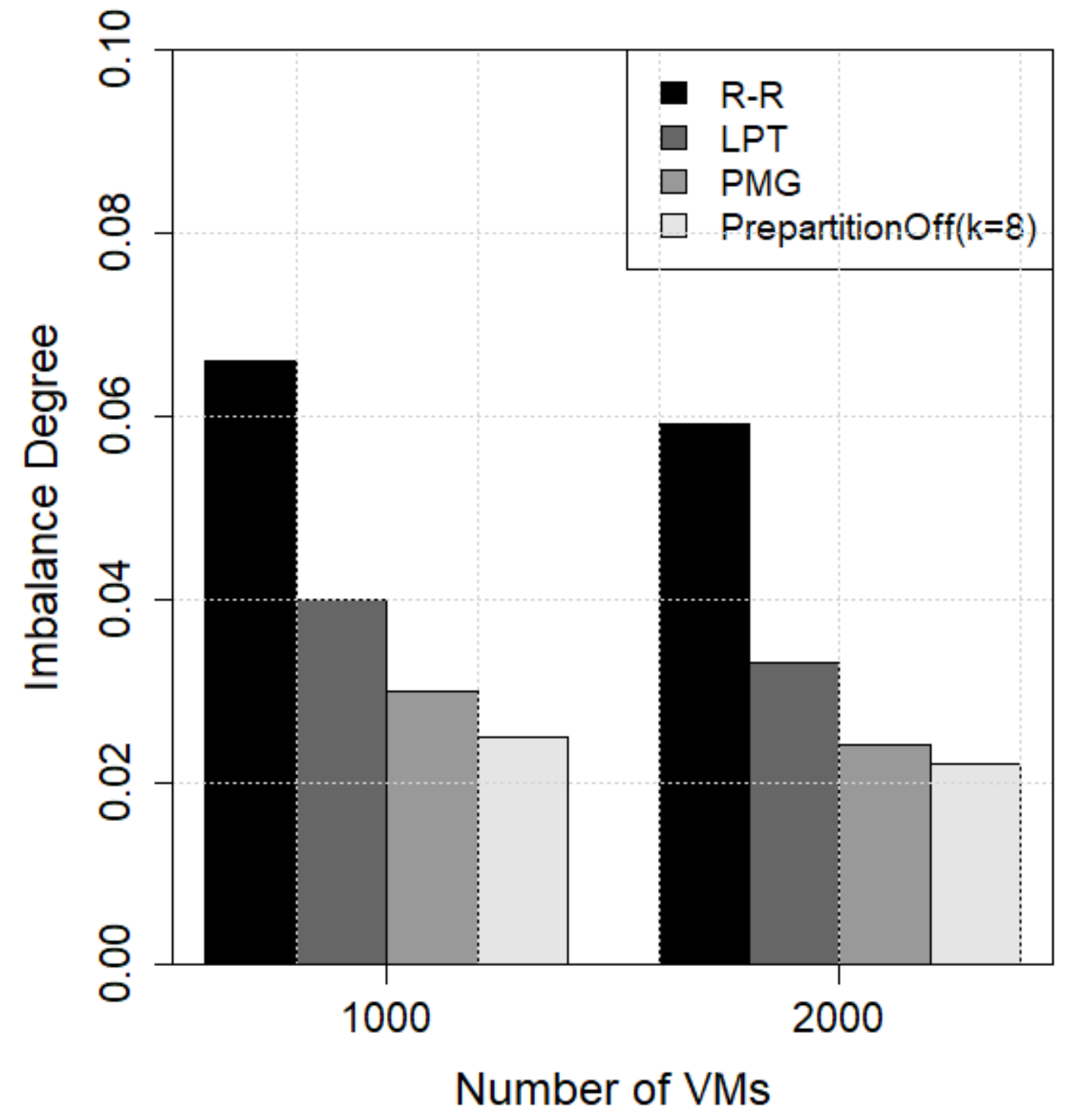}}}
	\subfloat[]{{\includegraphics[width=0.24\textwidth]{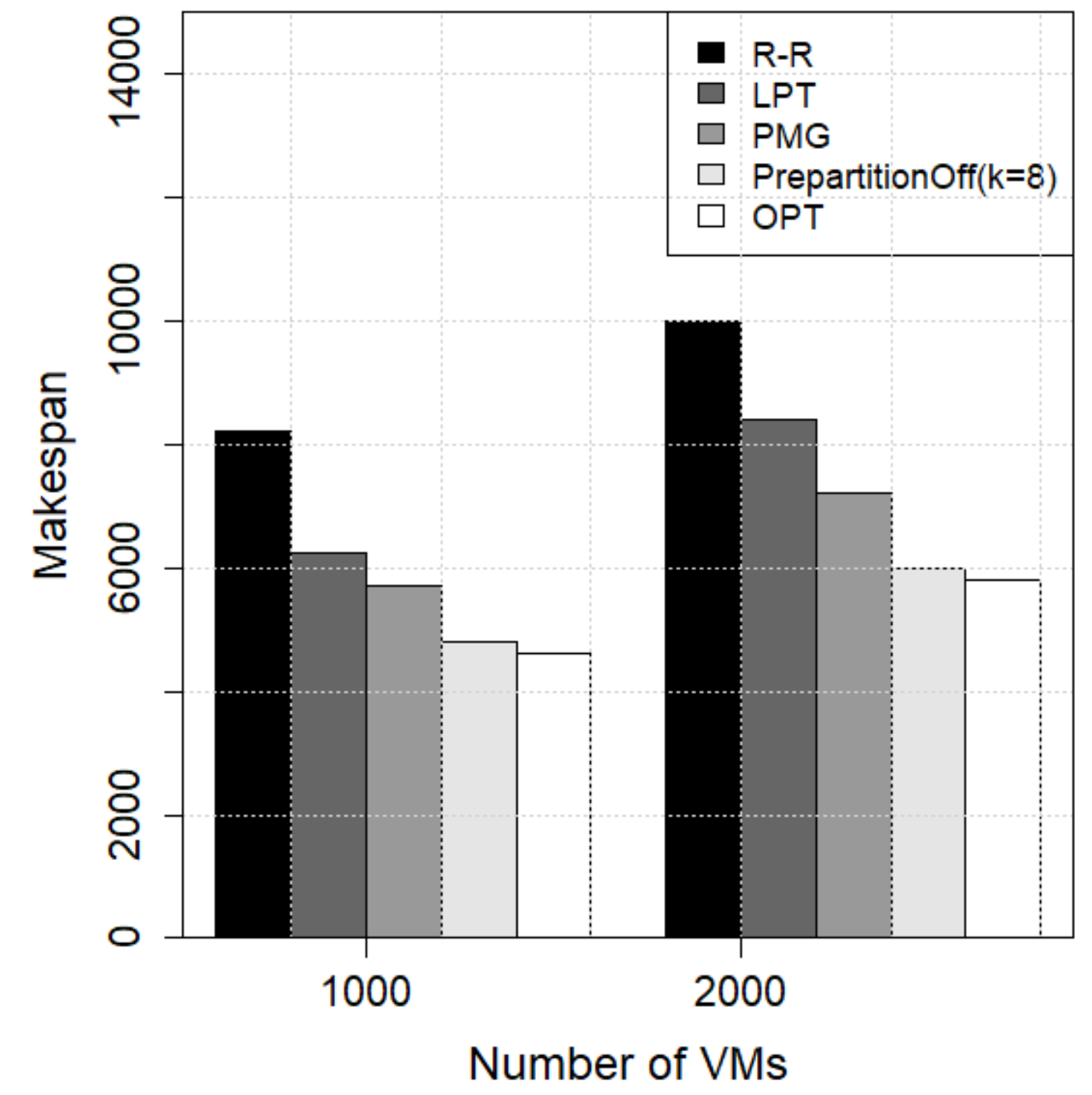}}}
	\subfloat[]{{\includegraphics[width=0.24\textwidth]{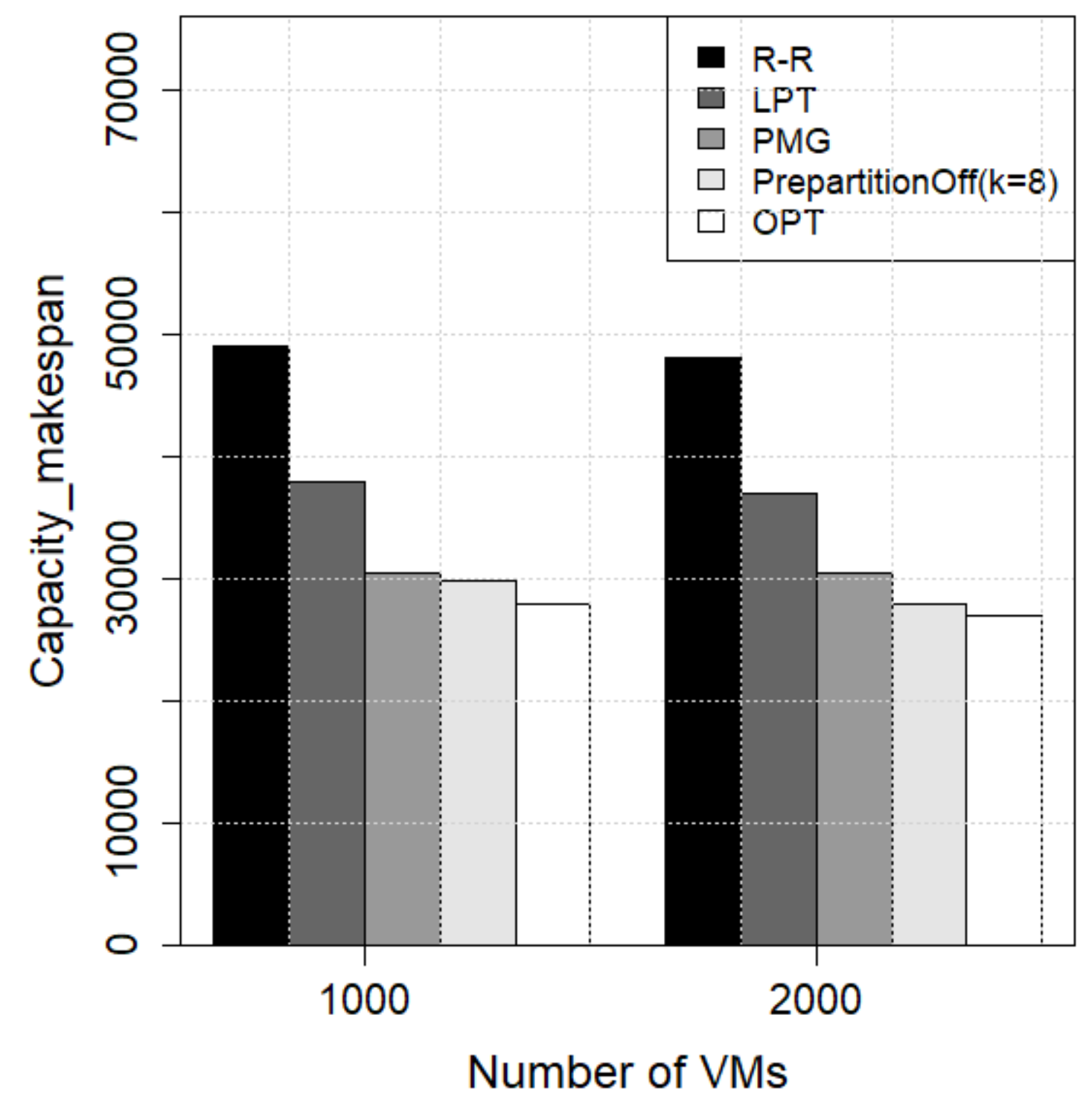}}}
	\caption{\color{black}The offline algorithm comparison of (a) average utilization with ESL trace; (b)  imbalance degree with ESL trace; (c) makespan with ESL trace; (d) Capacity$\_$makespan with ESL trace}
	\label{fig:offline-esl}
\end{figure*}

\noindent 1) \textbf{Round-Robin Algorithm (R-R)}: it is a load balancing scheduling algorithm by allocating the VM demands in turn to each PM that can offer
demanded resources.\\
2) \textbf{Longest Processing Time first (LPT)}: LPT is one of the best practices for offline scheduling algorithms without migration, which has an approximate ratio of 4/3. All the VM
demands are sorted by processing time in decreasing order firstly. Then demands are allocated to the PM with the smallest load in the sorted order. The smallest load indicates the smallest Capacity$\_$makespan among all the PMs.\\
3)\textbf{ Post Migration algorithm (PMG)}: PMG algorithm comes from the VMware DRS algorithm  \cite{IEEEhowto:Gulati},  which adopts migrations to achieve load balance regarding makespan. \color{blue}In the beginning, it allocates the demands the same way as LPT does. Here we replace makespan by Capacity$\_$makespan. 
Then the algorithm calculates the average Capacity$\_$makespan of all demands. 
In the PMG algorithm, the up-threshold and low-threshold are configured to achieve the load balancing effects, which are configured based on the average Capacity$\_$makespan and factor.
\color{black}In our experiments, we configure the factor as 0.1 (can be configured dynamically to meet the demands), which represents up-threshold is 1.1 times of the average Capacity$\_$makespan and the low-threshold is 0.9 multiples the average Capacity$\_$makespan.
The algorithm also maintains a migration list containing the VMs on the PMs with higher Capacity$\_$makespan value than the low-threshold. 
The VM migrations are triggered to make the PM to make the Capacity$\_$makespan smaller than the low-threshold. 	
Thereafter, the VMs in the migration list will be re-allocated to a PM with Capacity$\_$makespan smaller than the up-threshold. 
Migrating VMs to a new PM is triggered if the operation would not lead the Capacity$\_$makespan of the PM to be higher than the up-threshold. To be noted,  some VMs can be left in the list, thus finally the
algorithm allocates the left VMs to the PMs with the smallest Capacity$\_$makespan in sequence to balance the loads. 

VMs and PMs have the same configuration with Amazon EC2. The configurations are shown in Table \ref{tab:ECSspecification} and Table \ref{tab:PMspecification}, in which one unit of compute capacity equals to around
1.0 - 1.2 GHz 2007 Xeon or 2007 Opteron processors \cite{IEEEhowto:Amazon}.

\textit{Remarks:} We adopted the typical recommended VM types suggested by Amazon EC2. EC2 has a variety of VM types, and it classifies them as General Purpose, Compute Optimized (computational intensive VMs), Memory Optimized (memory-intensive VMs), Storage Optimized (storage-intensive VMs). Although
we adopted EC2 classification, our approach can still be extended to other classifications.


\subsubsection{\textbf{Replay with ESL Data Trace}}\label{Sec:FirstModel}
\color{black}	To reflect realistic data generation, we utilize the data derived from Experimental System Lab (ESL) \cite{IEEEhowto:ESL} that has been widely used for realistic data. \color{black}
The data with monthly records collected by the Linux cluster has characteristics that can be fitted into our model. In the log file, each line contains 18 elements where we only need parts of them, such as the requested ID,
start time, duration and the number of processors (capacity demands) in our simulation. Because the time slot length mentioned previously is set to 5 minutes, the units of the original data are converted from seconds to minutes.


Fig.~\ref{fig:offline-esl} shows the comparison of different algorithms in average utilization, imbalance degree, makespan and Capacity$\_$makespan. According to the results, we can observe that the PrepartitionOff algorithm can achieve better performance than other algorithms in four aspects. For average utilization, the PrepartitionOff algorithm is 10$\%$-20$\%$ higher than PMG, LPT, and Random-Robin (RR). The reason for different algorithms to have different average CPU utilization lies in that we consider heterogeneous
PMs and different algorithms may use the different number of total PMs. For makespan and Capacity$\_$makespan, the PrepartitionOff algorithm is 10$\%$-20$\%$ lower than PMG and LPI, 30$\%$-40$\%$ lower than Random-Robin (RR). And for imbalance degree, it is 30$\%$-40$\%$ lower than LPT.

\color{black}		\textbf{Observation 1}. As shown in the above performance evaluations, PMG is one of the best heuristic methods to balance loads, however, it can not assure a bounded or predefined target. \color{black}

\begin{figure*}[ht]
	
	\centering
	\subfloat[]{{\includegraphics[height=6cm, width=0.4\textwidth]{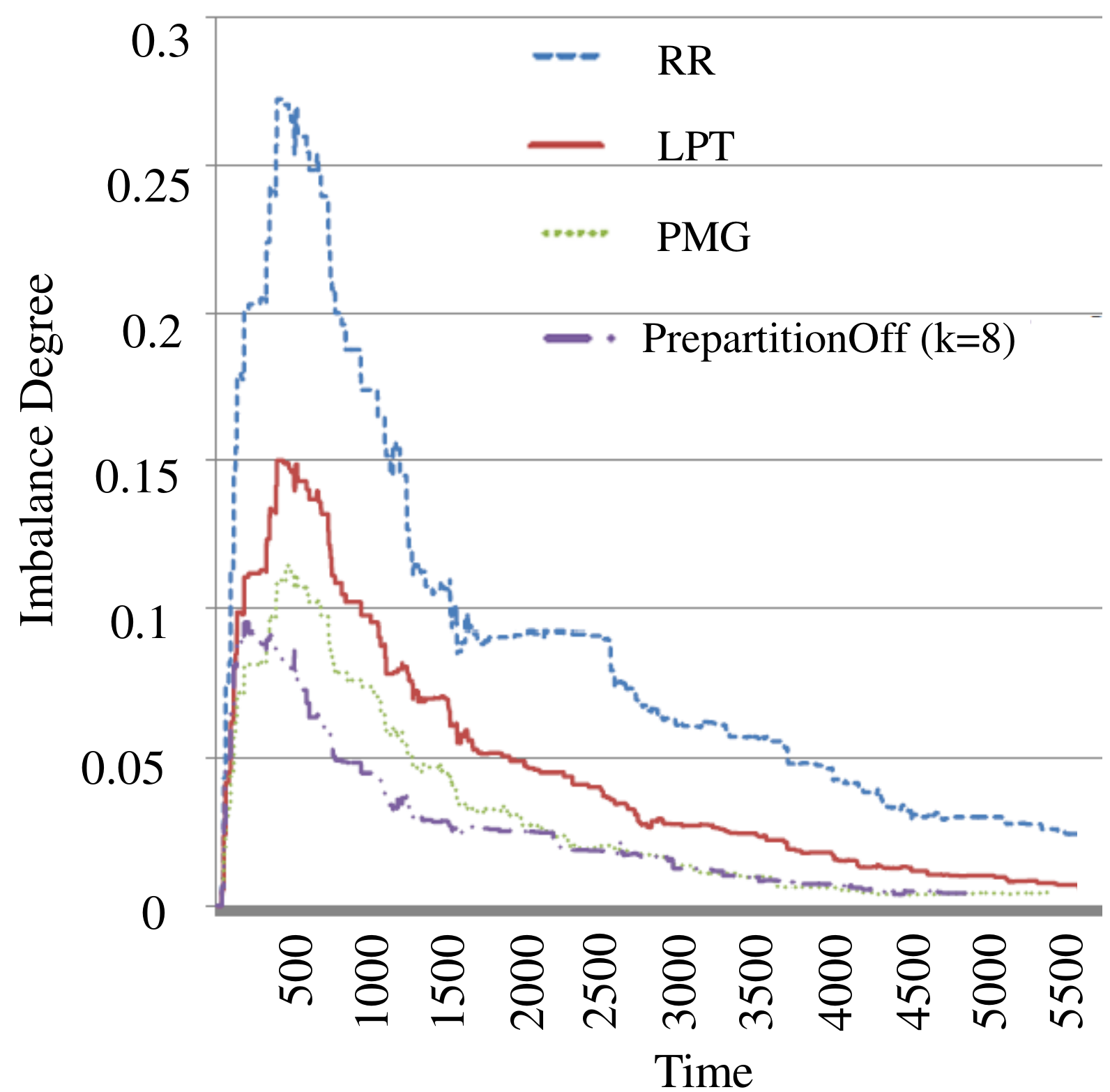}}}\hfill
	\subfloat[]{{\includegraphics[height=6cm, width=0.4\textwidth]{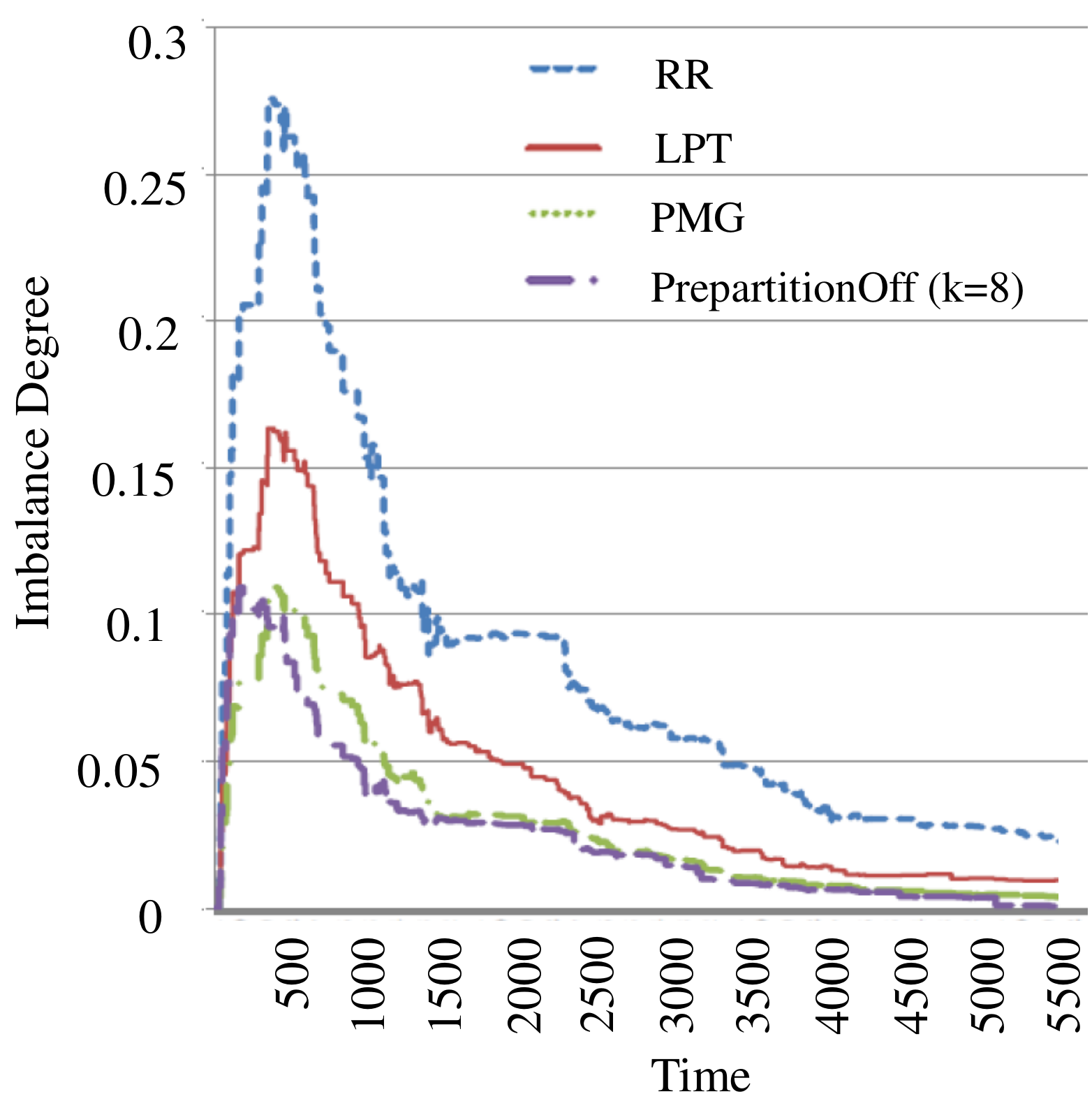}}}
	\caption{\color{blue}The consecutive imbalance degree under 1000 VMs among 4 different offline algorithms, where x-axis is for time and y-axis for imbalance degrees (a) 1000 VMs; (b) 2000 VMs.}
	\label{fig:offimb1000and200Vms}
\end{figure*}

\textbf{Observation 2}. PMG does not obtain the good performance as  PrepartitionOff in terms of average utilization, makespan and Capacity$\_$makespan, no matter what numbers of migration have been taken. \\ \\
The main reason is PrepartitionOff takes actions in a much more refined and desired scale by pre-partition based on reservation data while PMG is just a best-effort trial
by migration.
The reason is that PrepartitionOff is much more precise and desired with the aid of pre-partition while PMG is just a trial to balance load as much as possible.
To compare imbalance degree (IMD) change as time goes, we also did the tests about consecutive imbalance degree using 1000 and 2000 VMs among 4 different offline algorithms. In Fig ~\ref{fig:offimb1000and200Vms}, we provide the consecutive imbalance degree comparison for four algorithms in offline scheduling with 1000 VMs and 2000VMs respectively. In these two figures, the X-axis is for makespan and Y-axis is for imbalance degree. We can see that PrepartitionOff (with $k$=8) has the smallest makespan and smallest imbalance degree most of the time during tests, except for the initial period. Notice that the value of $k$ can be set differently, here we just present the results for $k$=8.

%
%

\subsubsection{\textbf{Results Comparison by Synthetic Data} }
We configure the time slot to be 5 minutes as mentioned before, so an hour has 12 slots and a day has 288 slots. All requests are subject to Normal distribution with mean $\mu$ as 864 (three days) and standard deviation $\delta$  as 288 (one day) respectively. After requests are generated in this way, we start the simulator to simulate the
scheduling effects of different algorithms and comparison results are collected.
\color{black}	For data collection, first we set $k$ of PrepartitionOff algorithm as 4 (we configure the value as 4 because in previous research \cite{Tian2014a}, this value has been validated to be an effective value to improve performance). \color{black}And the different types of VM are with equal probabilities. We also vary the number of VMs from 100, 200, 400 and 800 to analyze the trend. Each data set is an average of 10-runs.
\begin{figure*}[ht]
	\centering
	\subfloat[]{{\includegraphics[width=0.24\textwidth]{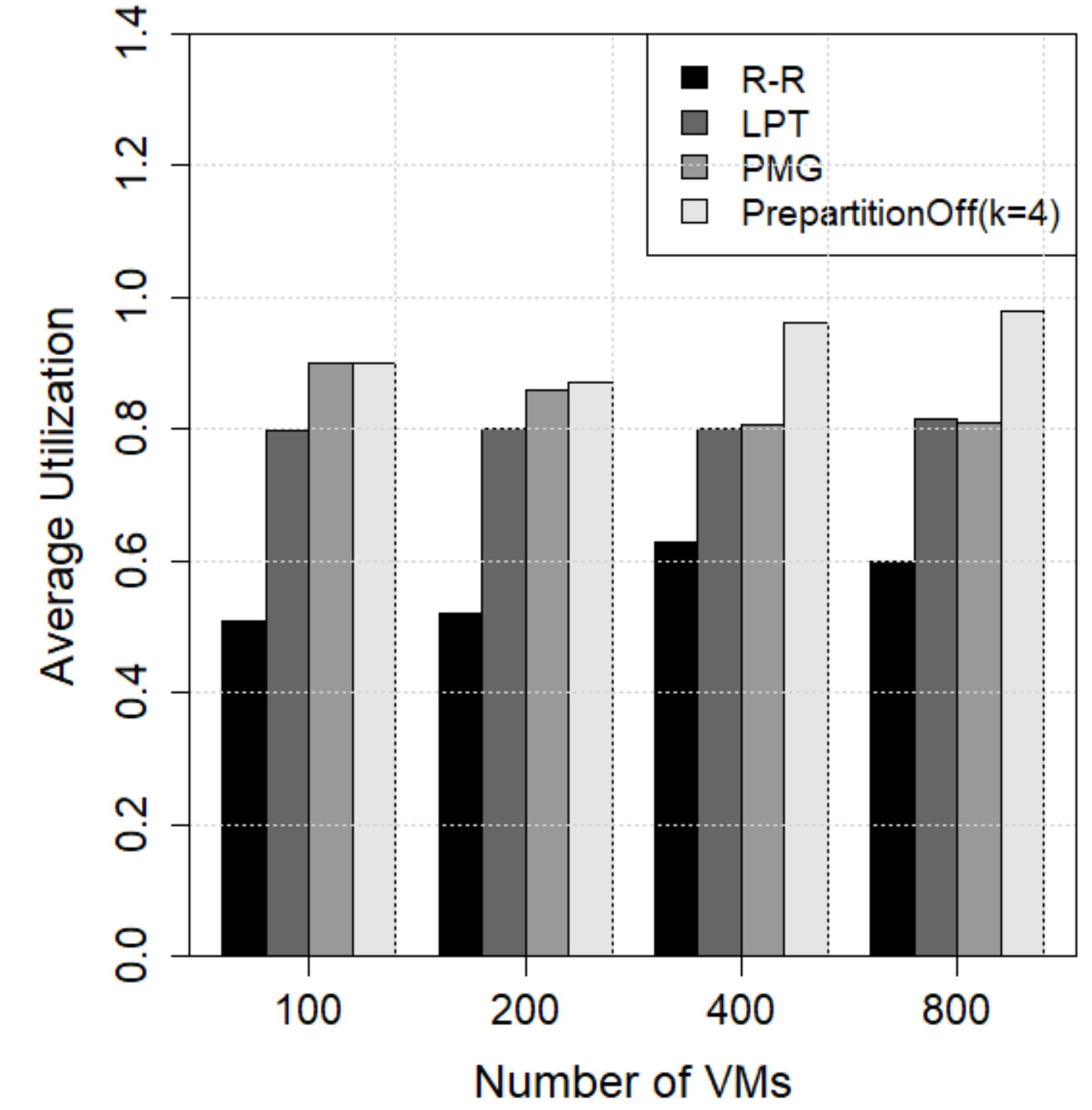}}}\hfill
	\subfloat[]{{\includegraphics[width=0.24\textwidth]{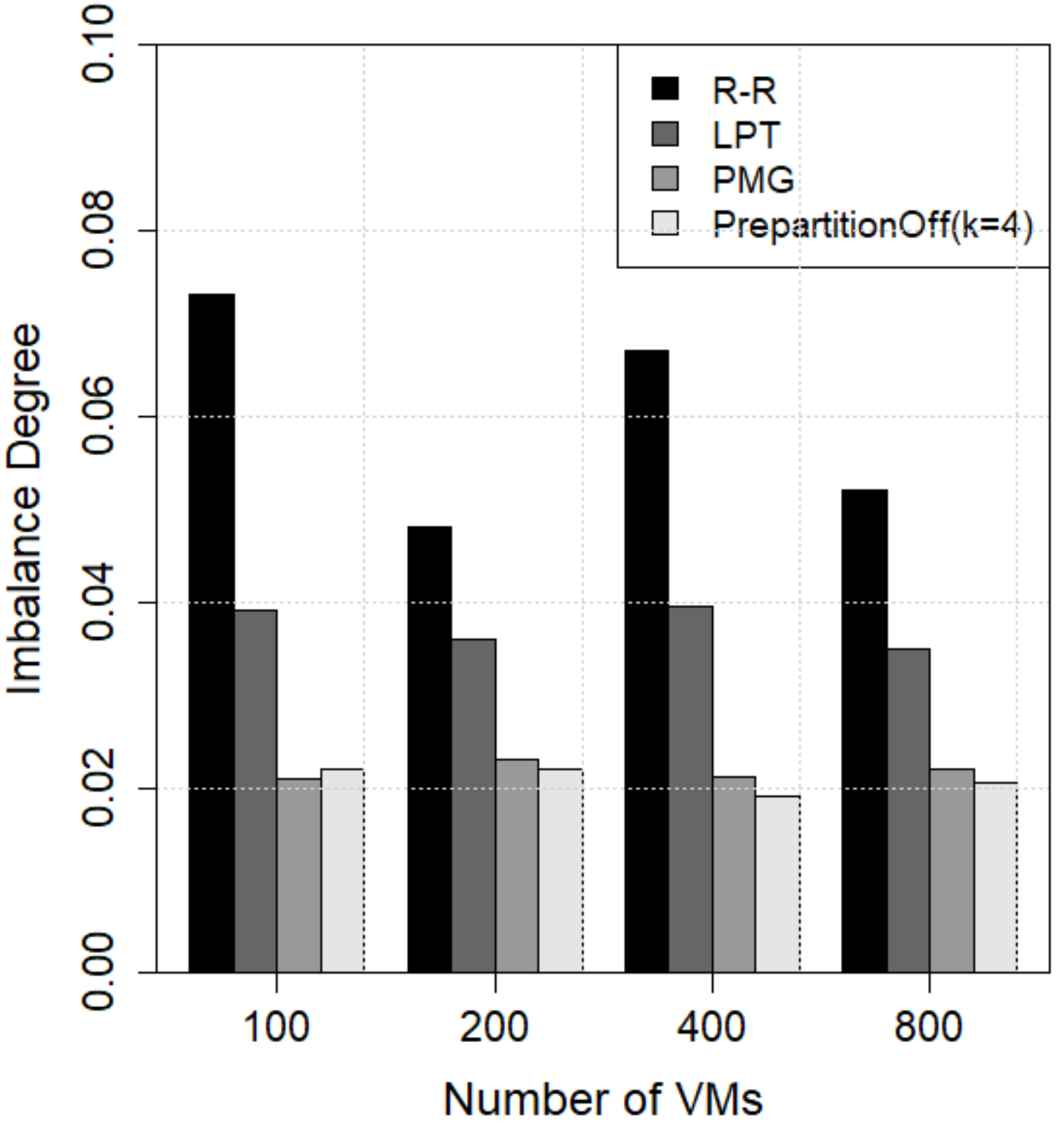}}}
	\subfloat[]{{\includegraphics[width=0.24\textwidth]{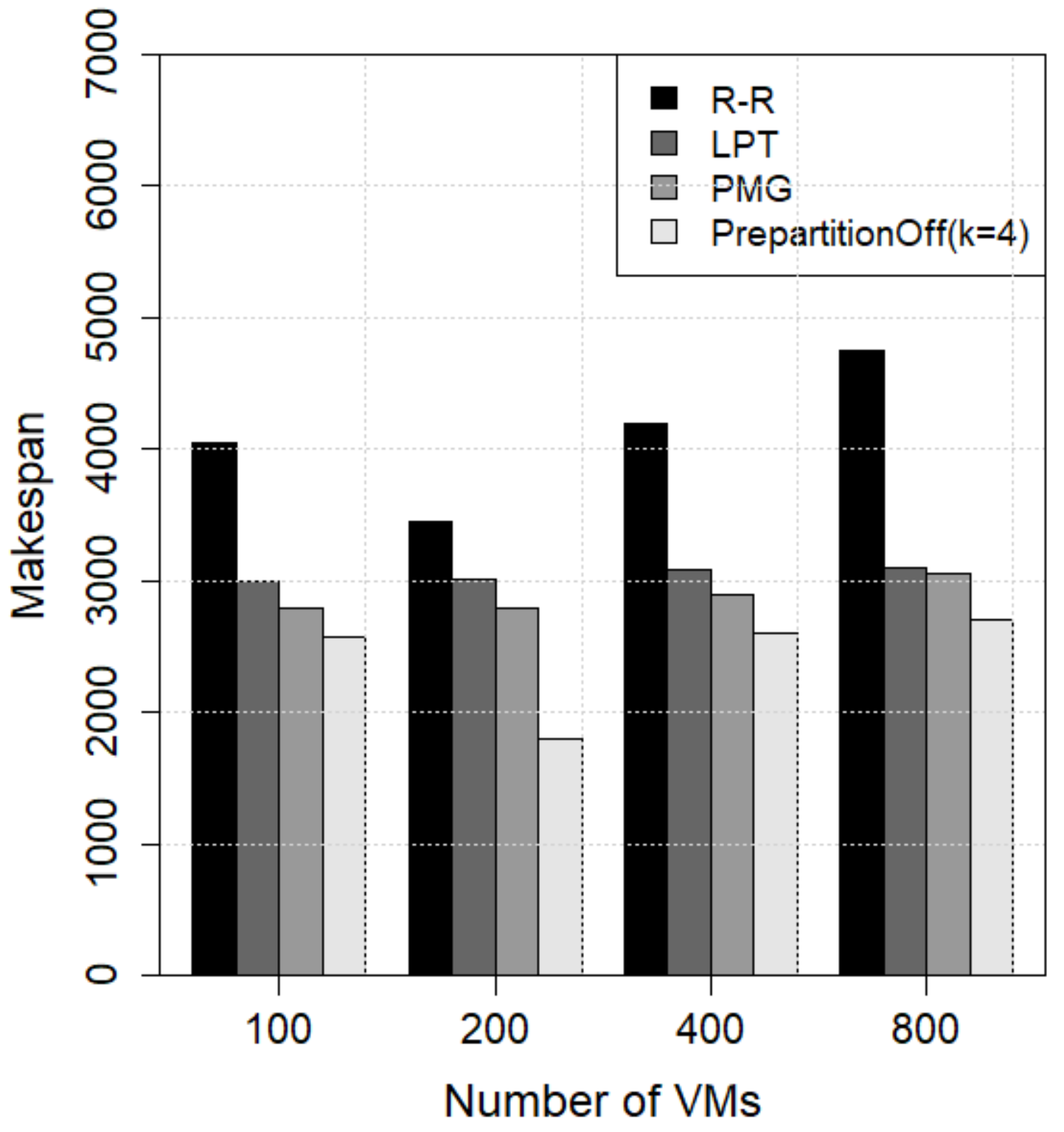}}}
	\subfloat[]{{\includegraphics[width=0.24\textwidth]{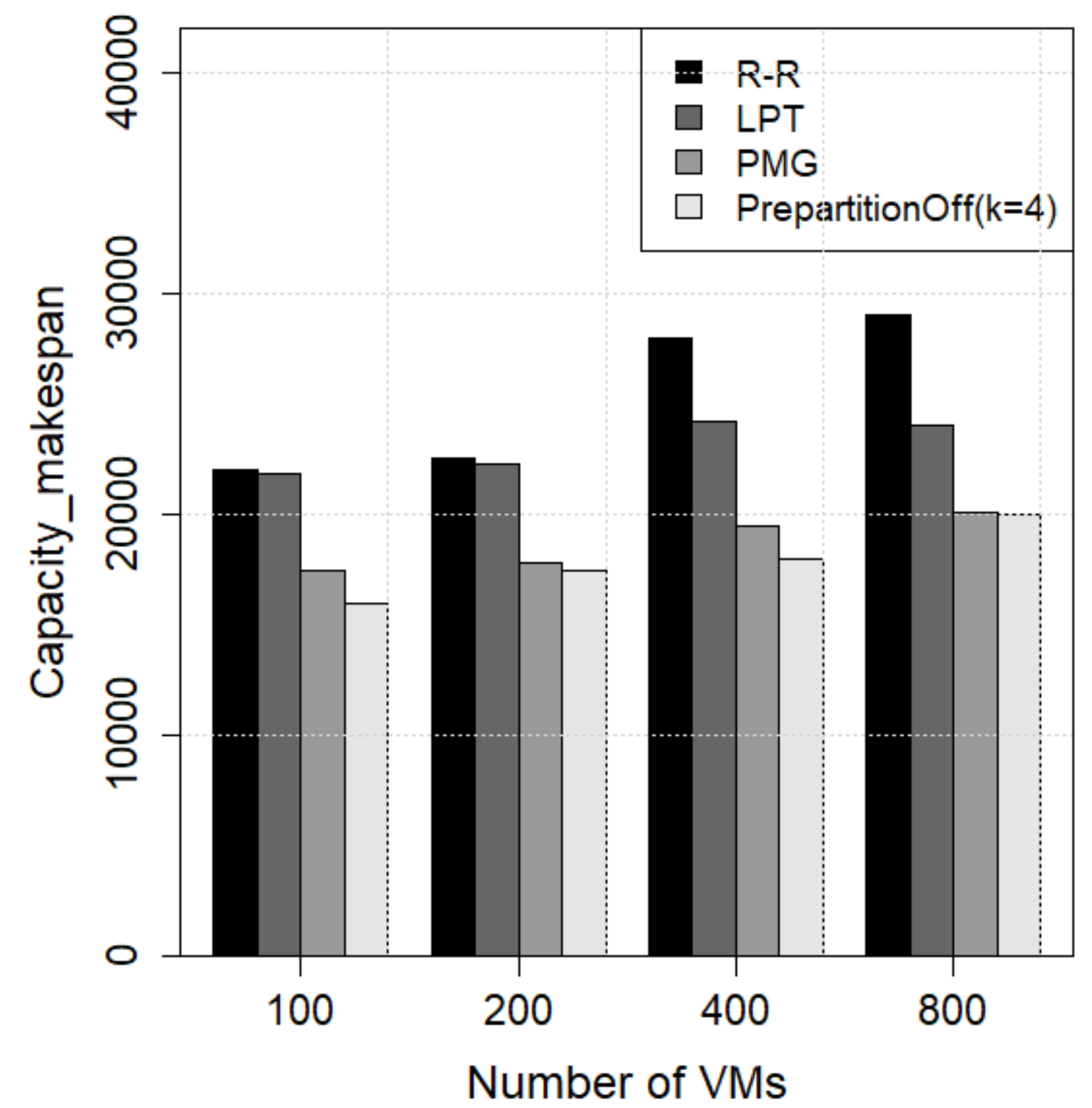}}}
	\caption{The comparison of offline algorithm with (a) average utilization; (b) imbalance degree; (c) makespan; (d) capacity\_makespan with Normal distribution }
	\label{fig:offlinenormal}
\end{figure*}



Fig. \ref{fig:offlinenormal} displays the comparison of different algorithms in average utilization,
makespan and Capacity$\_$makespan. From these figures, we can know that the PrepartitionOff algorithm is 10$\%$-20$\%$ higher than PMG and LPT for average utilization,
40$\%$-50$\%$ higher than Random-Robin (RR). As for makespan and Capacity$\_$Makespan, the PrepartitionOff algorithm is 8$\%$-13$\%$ lower than PMG and LPT, 40$\%$-50$\%$ lower than Round-Robin (RR).
\color{blue}We also note that the PMG algorithm can improve the performance of the LPT algorithm as it configures up-threshold and low-threshold based on Capacity\_makespan value. \color{black}LPT algorithm is better than the RR algorithm. Similar results are observed for the comparison of makespan.


\subsection{PrepartitionOn1 Algorithm}
We demonstrate the simulation results of the PrepartitionOn1 algorithm and the other three algorithms in this section.
\color{black}All VM requests are generated by following normal distribution, and Random, RR, \color{blue}Online Resource Scheduling Algorithm (OLRSA) \cite{IEEEhowto:Xu} that has a good competitive ratio ($2-\frac{1}{m}$, where $m$ is the number of PMs) \color{black}for online algorithm has been implemented to compare with PrepartitionOn1. 
OLRSA calculates the Capacity$\_$makespan of all the PMs and sorts PM by Capacity$\_$makespan in descending order, which assigns the VM request to the PM with the lowest Capacity$\_$makespan and required resources.\color{black}

\begin{figure*}[ht]
	\centering
	\subfloat[]{{\includegraphics[width=0.24\textwidth]{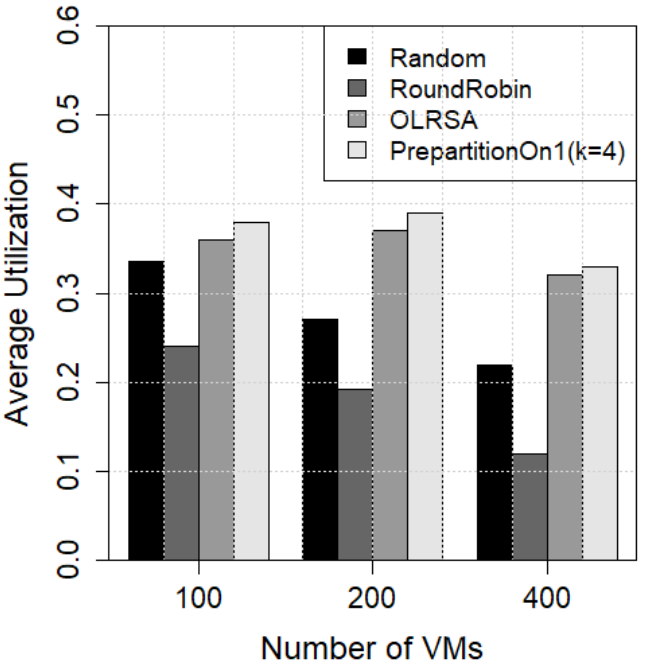}}}\hfill
	\subfloat[]{{\includegraphics[width=0.24\textwidth]{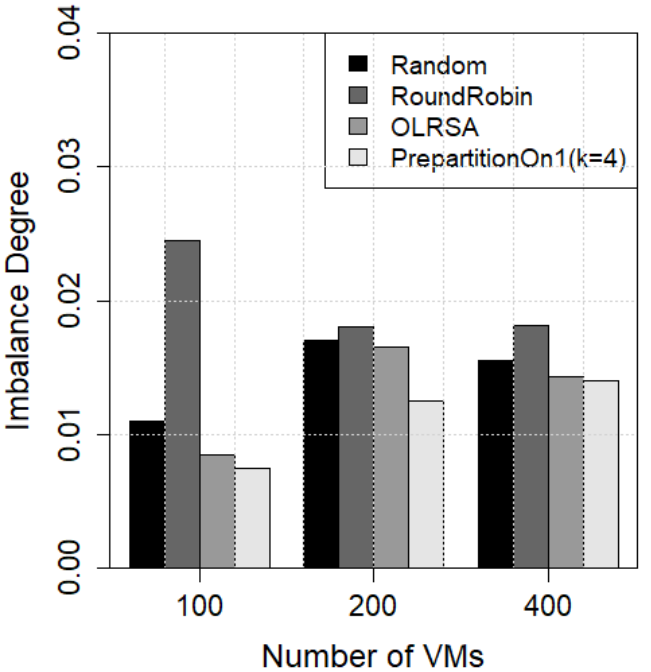}}}
	\subfloat[]{{\includegraphics[width=0.24\textwidth]{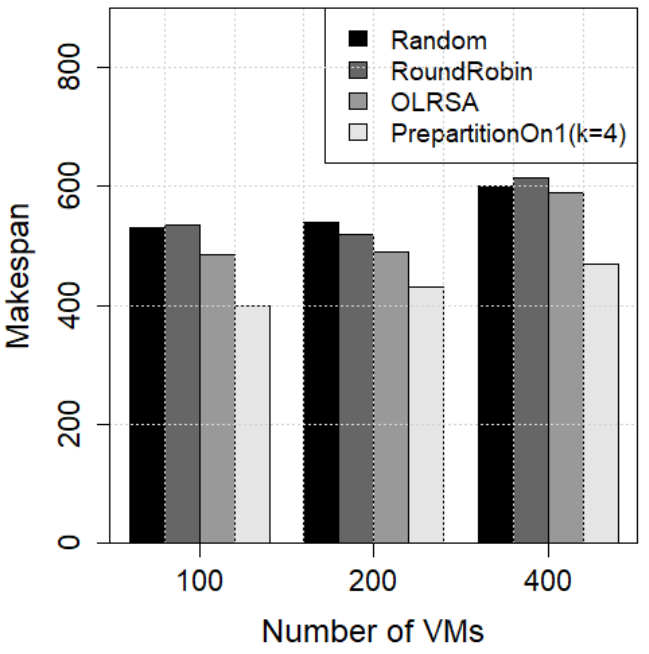}}}
	\subfloat[]{{\includegraphics[width=0.24\textwidth]{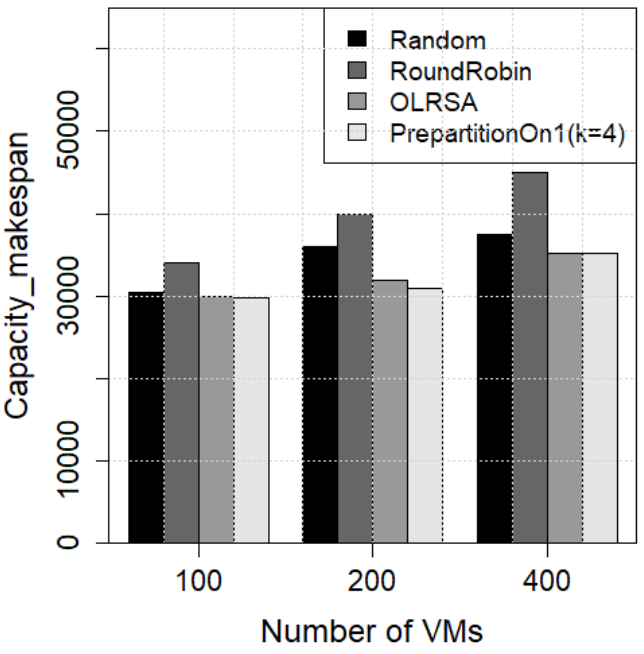}}}
	\caption{The comparison of online algorithm with (a) average utilization; (b) imbalance degree; (c) makespan; (d) capacity\_makespan with ESL trace}
	\label{fig:onlineesl}
\end{figure*}

\subsubsection{\textbf{Replay with ESL Data Trace}}

\begin{figure*}[!ht]
	\begin{center}
	\end{center}
\end{figure*}
\color{black}	The ESL dataset aforementioned is also used in the experiments.
\color{black}Fig.~\ref{fig:onlineesl} illustrates the comparisons of the average utilization, imbalance degree, makespan, Capacity$\_$makespan. According to these figures, we can see that PrepartitionOn1 demonstrates the highest average utilization, the lowest imbalance degree, and the lowest makespan. As for Capacity$\_$makespan, OLRSA
has been proved much better performance compared with random and round-robin algorithms, and PrepartitionOn1 still improves 10$\%$-15$\%$ in average utilization,
20$\%$-30$\%$ in imbalance degree, and 5$\%$ to 20$\%$ in makespan than OLRSA.
\subsubsection{\textbf{Results Comparison by Synthetic Data}}

\begin{figure*}[ht]
	\centering
	\subfloat[]{{\includegraphics[width=0.24\textwidth]{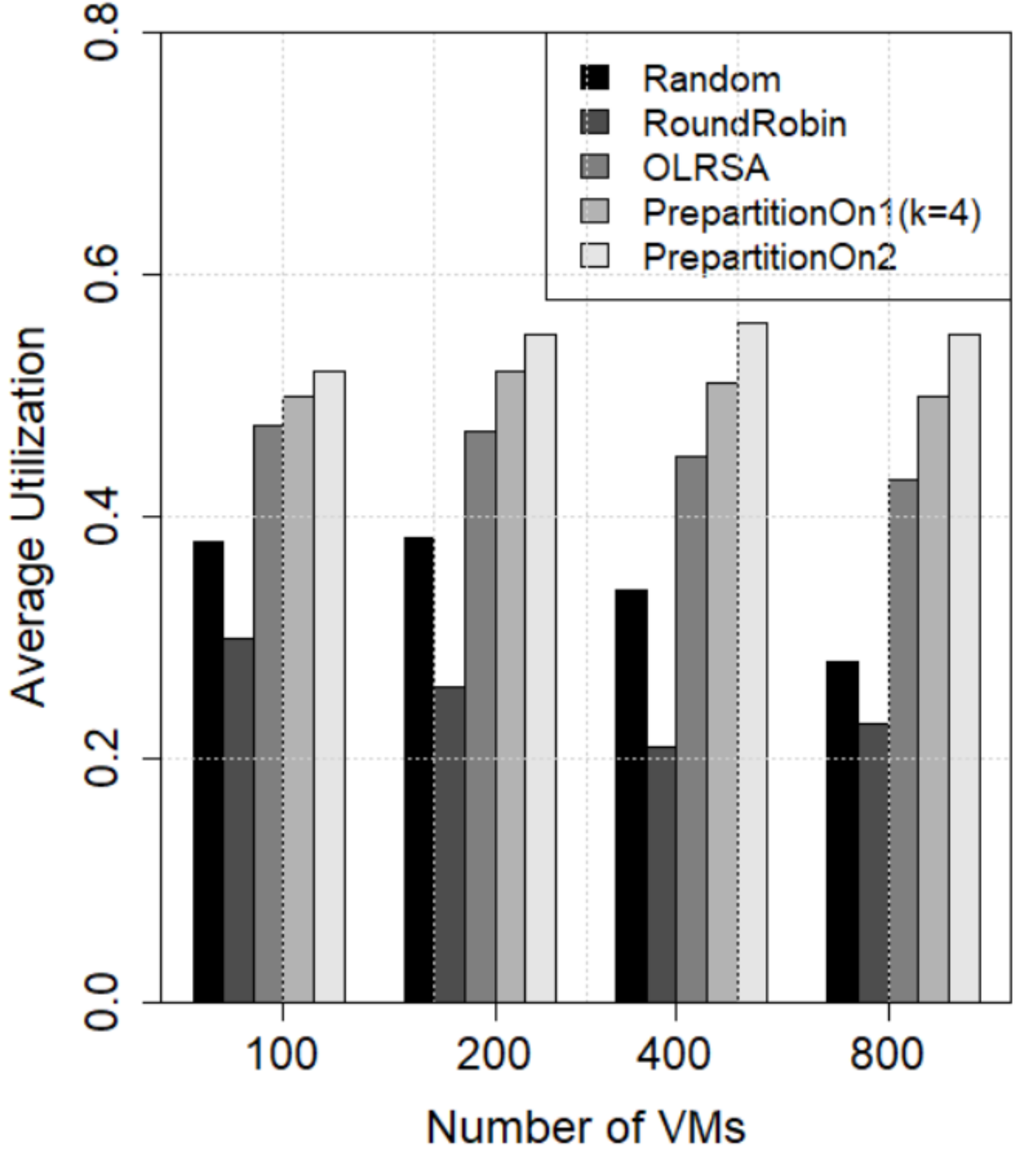}}}\hfill
	\subfloat[]{{\includegraphics[width=0.24\textwidth]{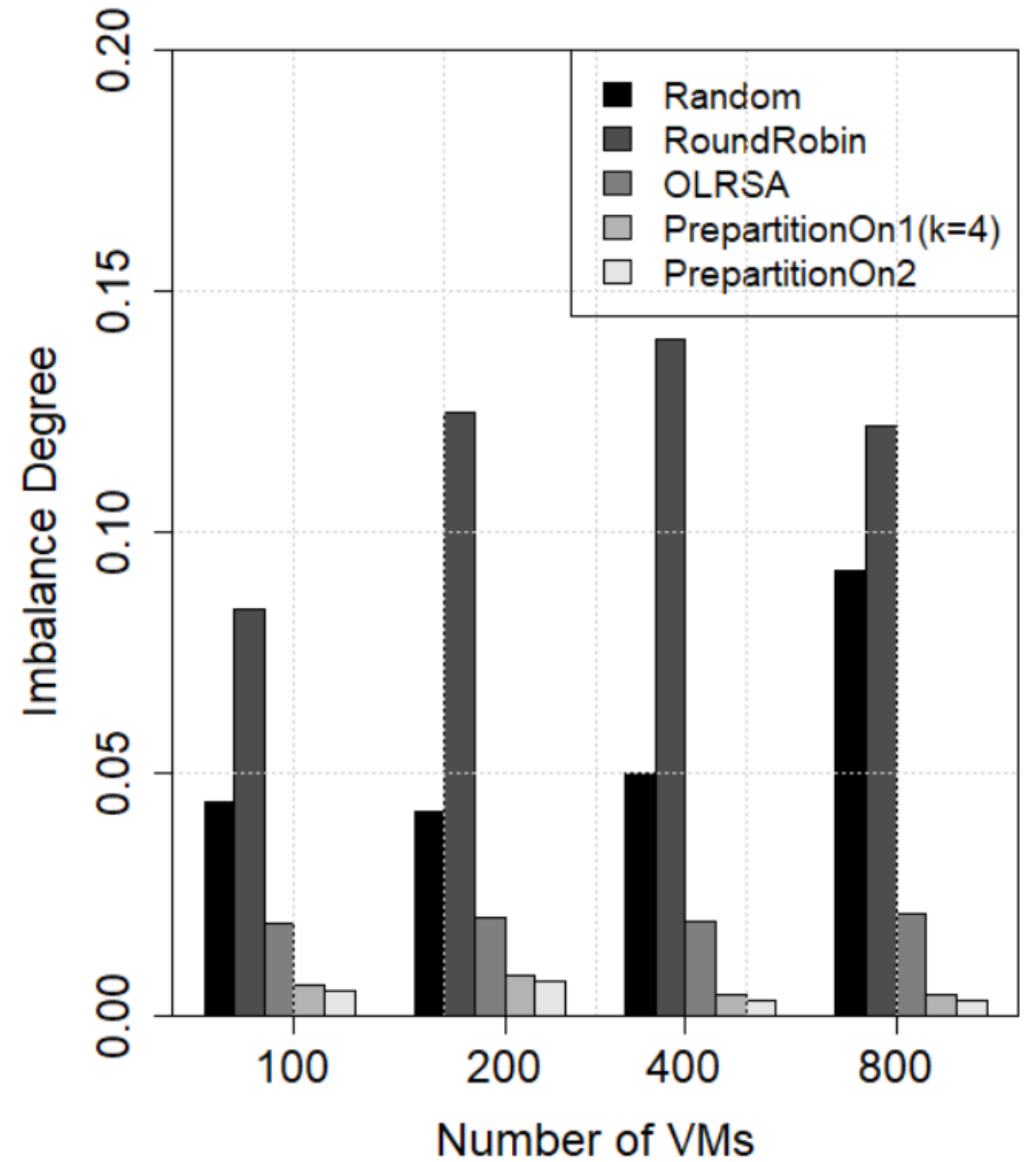}}}
	\subfloat[]{{\includegraphics[width=0.24\textwidth]{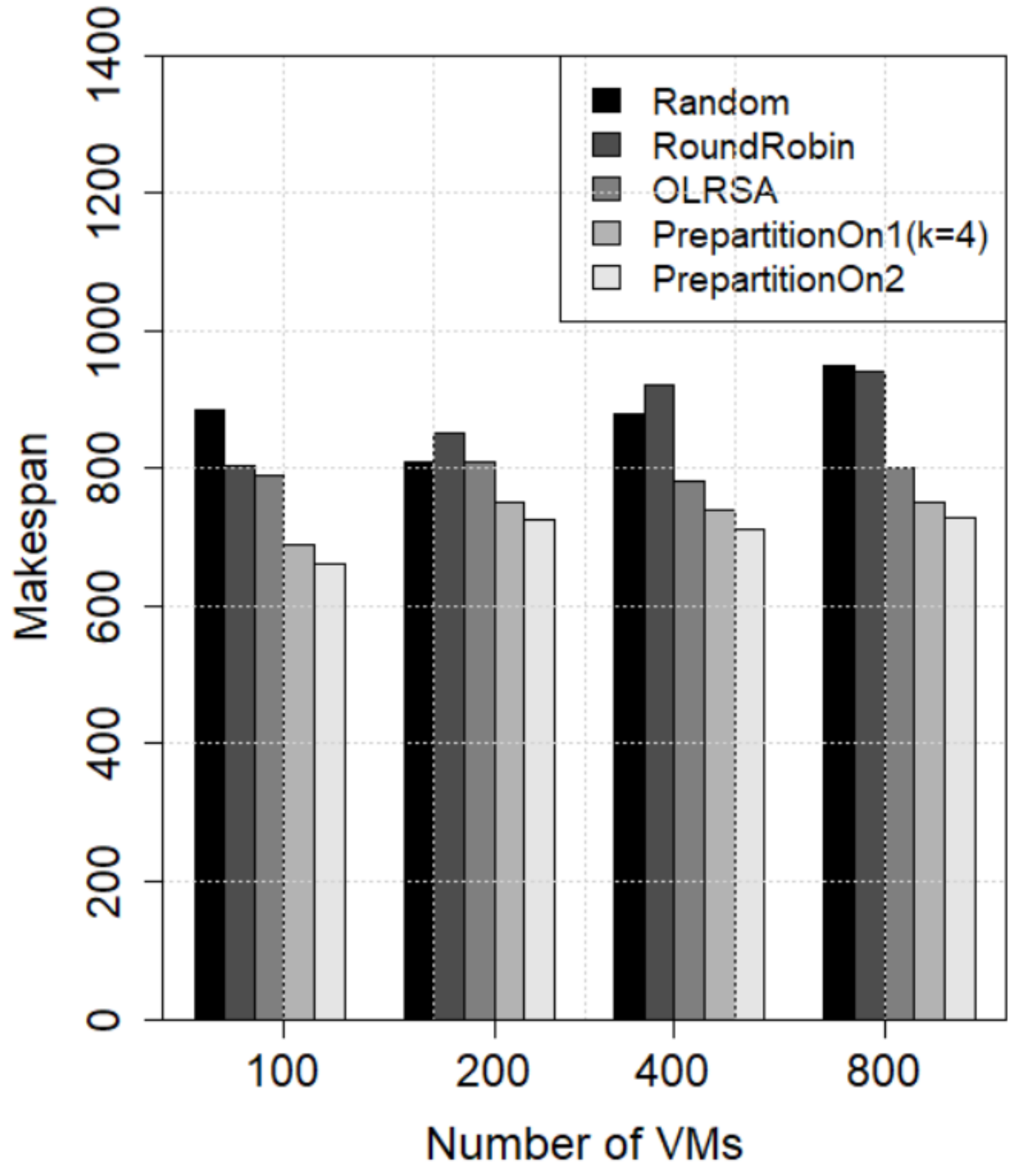}}}
	\subfloat[]{{\includegraphics[width=0.24\textwidth]{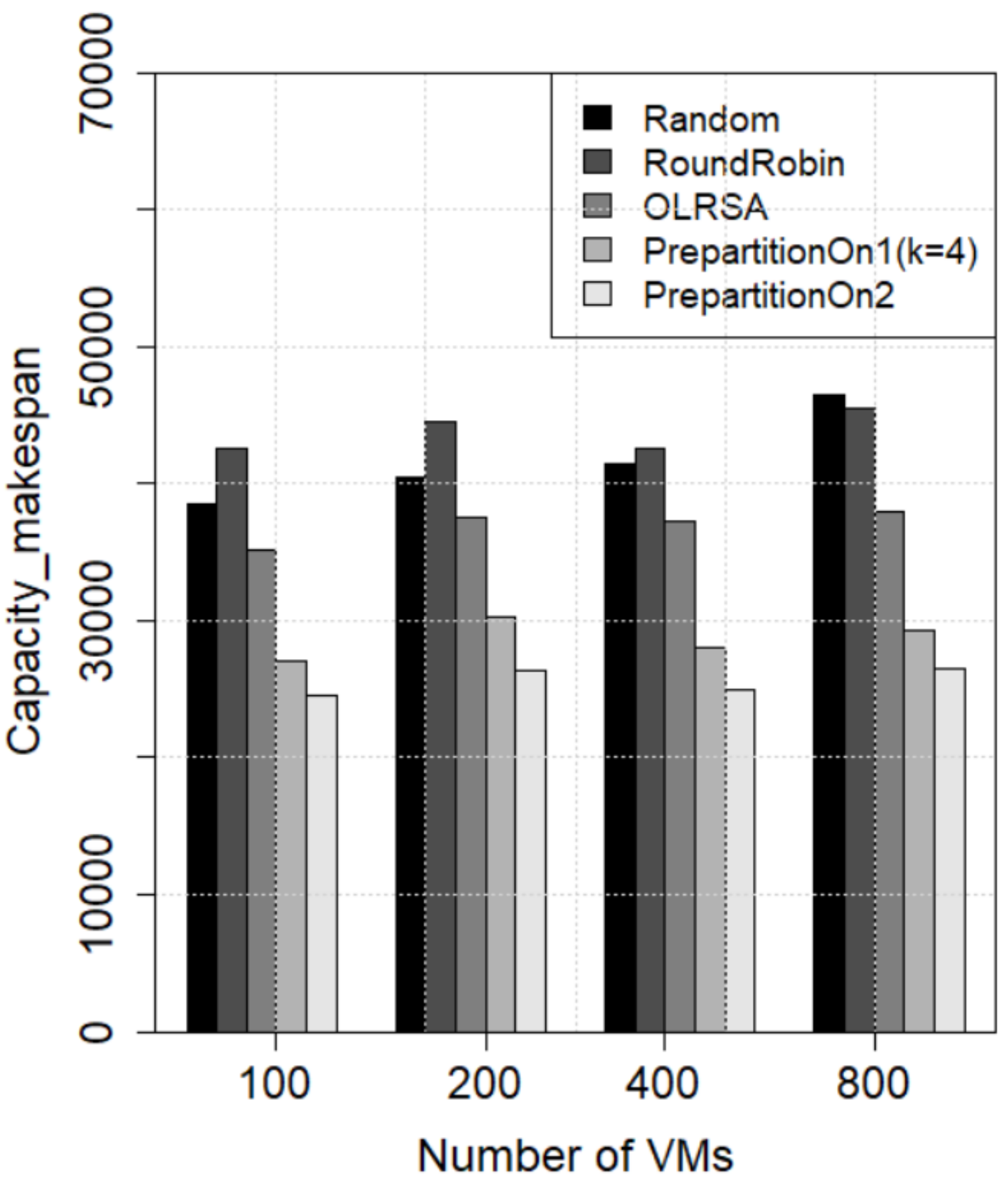}}}
	\caption{\color{black}The online algorithm comparison of (a) average utilization; (b) imbalance degree; (c) makespan; (d) capacity\_makespan with normal distribution}
	\label{fig:onnormal}
\end{figure*}



The requests are configured as same as in Section 5.1 based on the normal distribution.
We set that VMs with different types have equal probabilities, and we modify the requests generation approach to produce different sizes of requests to trace the tendency. From Fig.~\ref{fig:onnormal}, we can see that PrepartitionOn1 has better performance in average utilization, imbalance degree, makespan and Capacity$\_$makespan. Comparing with OLRSA, PrepartitionOn1 still improves about 10$\%$
in average utilization, 30$\%$-40$\%$ in imbalance degree, 10$\%$-20$\%$ in makespan, as well as 10$\%$-20$\%$ in
Capacity$\_$makespan.

LPT is one kind of the best methods for offline load balance algorithms without migration, which has an approximate ratio of 4/3. So we suggest setting the
$k$ value as 4, which can obtain an approximate ratio as 1+ $1/k$= 5/4. Under this configuration, a better approximate ratio could be obtained. With higher $k$
value, better load balancing effects could be achieved. While there exist tradeoffs between load balancing effect and time cost. For online load balance algorithms, we also suggest setting the $k$ value as 4, and cloud service providers could reconfigure that value to be higher as suitable as the load balancing effects they desired. 

\color{blue} Let us consider that we have $m=100$ PMs and the $k$ value is set as 4, then according to the analysis in~\cite{IEEEhowto:Xu}, the complexity ratio of OLRSA is $2-\frac{1}{m}=2-\frac{1}{100}$, and the complexity ratio of PrepartitionOn1 is $1+\frac{1}{k}-\frac{1}{mk} = 1+\frac{1}{4}-\frac{1}{400}$ based on Section 4.2. This proves that  PrepartitionOn1 can achieve better performance than OLRSA theoretically. 
\color{black}

\subsection{PrepartitionOn2 Algorithm}

In this part, we display the simulation results of the PrepartitionOn2 algorithm and the other three algorithms. Random, Round-Robin, Online Resource Scheduling Algorithm (OLRSA) \cite{IEEEhowto:Xu} and PrepartitionOn2 Algorithm are implemented for comparison.

\subsubsection{Replay with ESL Data Trace and Synthetic Trace}

\begin{figure*}[ht]
	\centering
	\subfloat[]{{\includegraphics[width=0.24\textwidth]{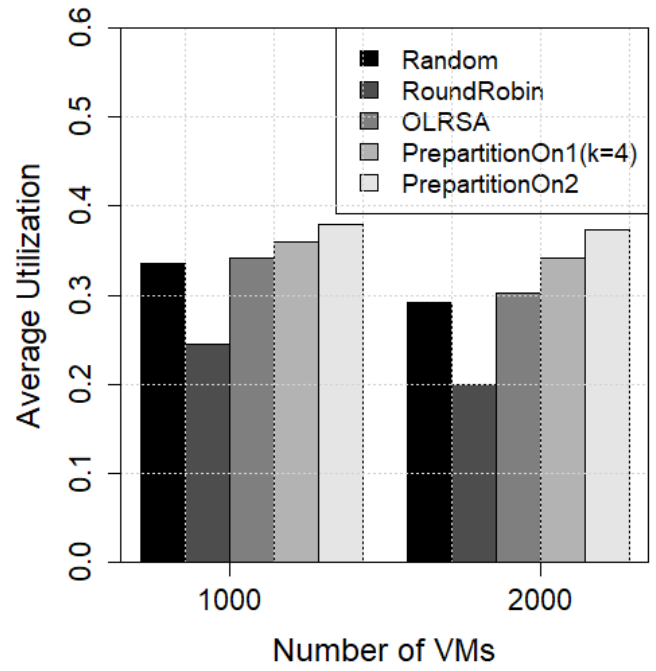}}}\hfill
	\subfloat[]{{\includegraphics[width=0.24\textwidth]{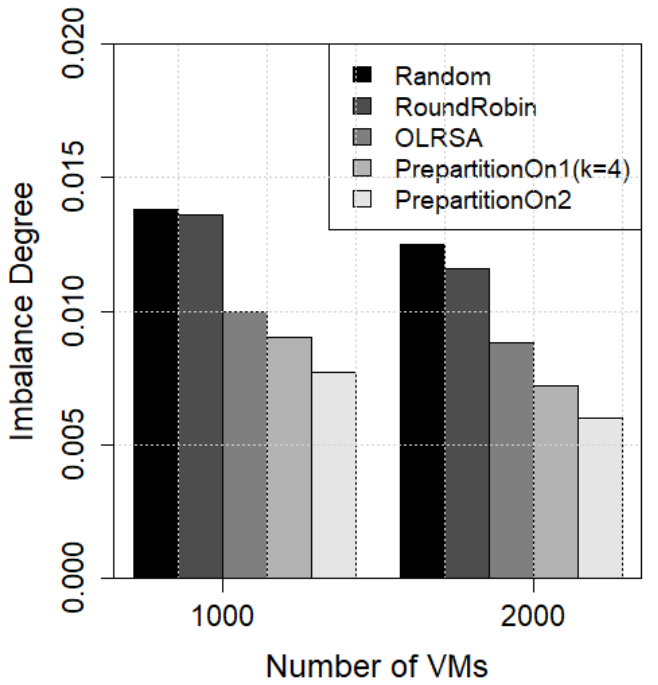}}}
	\subfloat[]{{\includegraphics[width=0.24\textwidth]{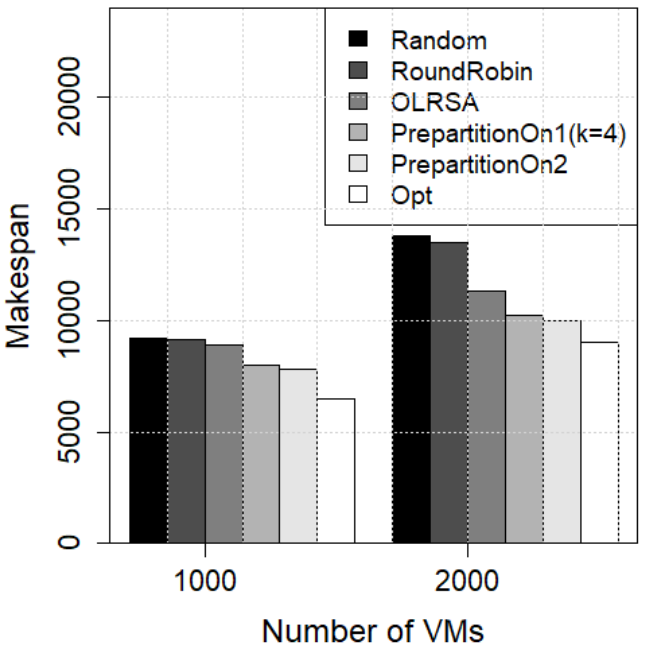}}}
	\subfloat[]{{\includegraphics[width=0.24\textwidth]{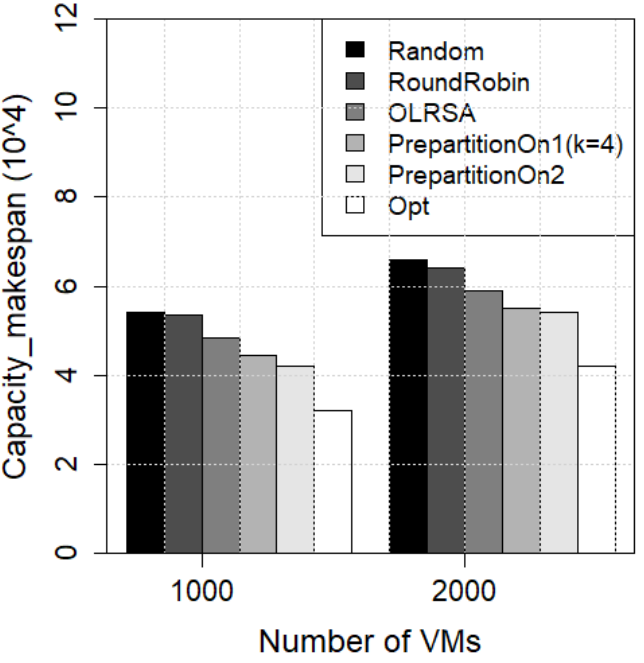}}}
	\caption{The online algorithms and Prepartition algorithm comparison of (a) average utilization; (b) imbalance degree; (c) makespan; (d) capacity\_makespan with ESL distribution}
	\label{fig:onlinenormal}
\end{figure*}

\color{black}We still use the log data from ESL and normal distribution for experiments. \color{black}Fig. ~\ref{fig:onnormal} and Fig.~\ref{fig:onlinenormal} illustrate the comparisons of the average utilization, imbalance degree, makespan, Capacity$\_$makespan between PrepartitionOn2 and other online algorithms and the results show that PrepartitionOn2 performances best in terms of mentioned metrics.

\begin{figure*}[ht]
	\centering
	\subfloat[]{{\includegraphics[height=6cm, width=0.4\textwidth]{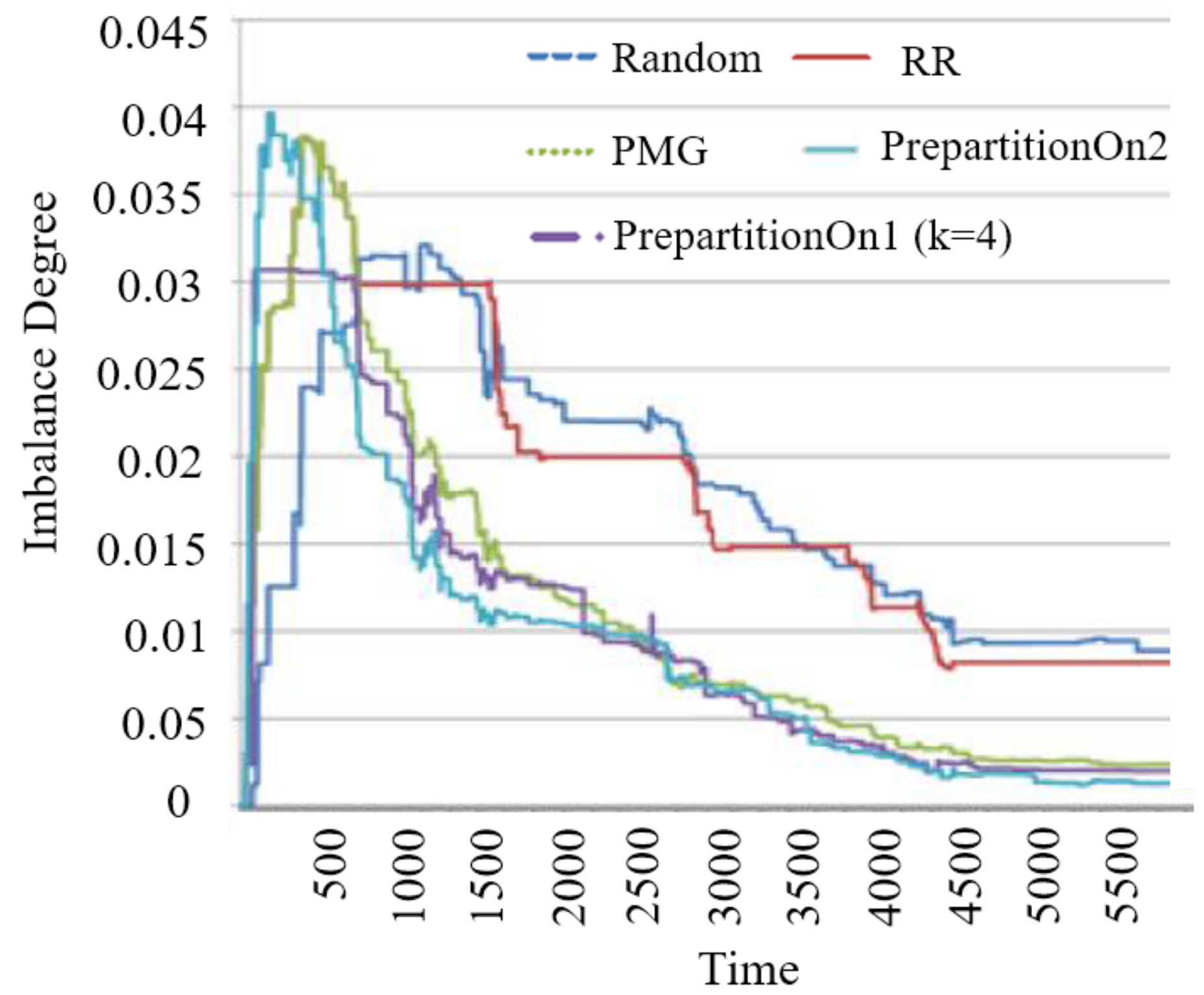}}}\hfill
	\subfloat[]{{\includegraphics[height=6cm,
			width=0.44\textwidth]{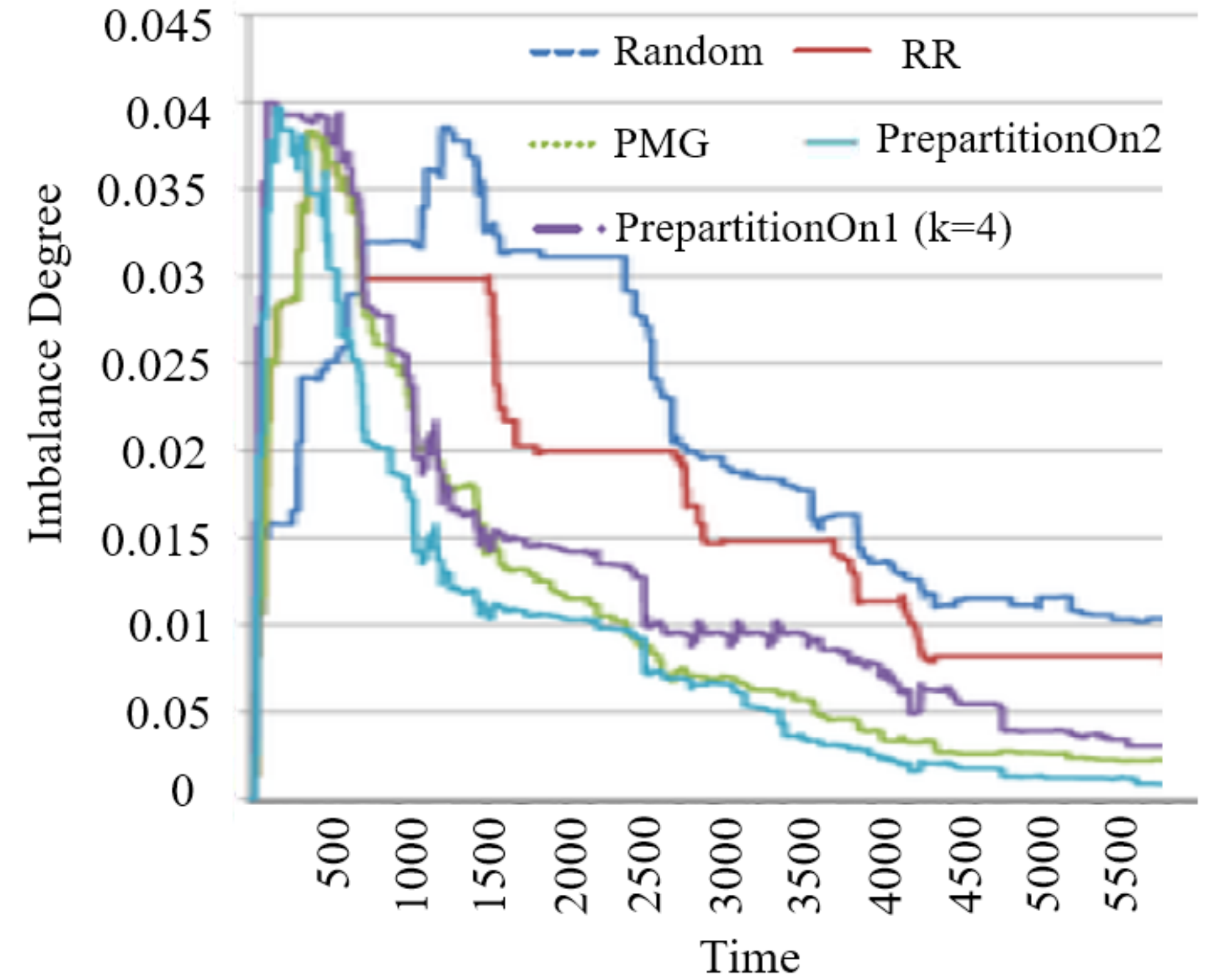}}}
	\caption{\color{blue}The consecutive imbalance degree under 1000 VMs among 5 different online algorithms, where x-axis is for time and y-axis for imbalance degrees (a) 1000 VMs; (b) 2000 VMs.}
	\label{fig:onlineimb1000and2000Vms}
\end{figure*}

In Fig.~\ref{fig:onlineimb1000and2000Vms}, we provide the consecutive imbalance degree comparison for four algorithms in online scheduling with 1000 VMs and 2000VMs respectively. In these two figures, the X-axis is for makespan and Y-axis is for imbalance degree. We can see that PrepartitionOn2 has the smallest makespan and smallest imbalance degree most of the time during tests.

The large $k$ values may bring side effects since it will need more partitions. In Fig. 10, we compare the time costs (simulated with
ESL data and the time unit is millisecond) under different partition value $k$, PrepartitionOn1 algorithm with $k=3$ takes about 10\% less running time than that
with $k=4$, and $k=2$ takes 15\% less running time than that with $k=4$. A larger $k$ value will lead to a better load balance with a longer process time. We also observe that a larger $k$ value will induce a lower Capacity$\_$makespan value. Similarly, with a larger $k$ value, larger average utilization, lower imbalance degree, and makespan are obtained. 

\begin{center}
	\includegraphics [width=.35\textwidth, angle=-0] {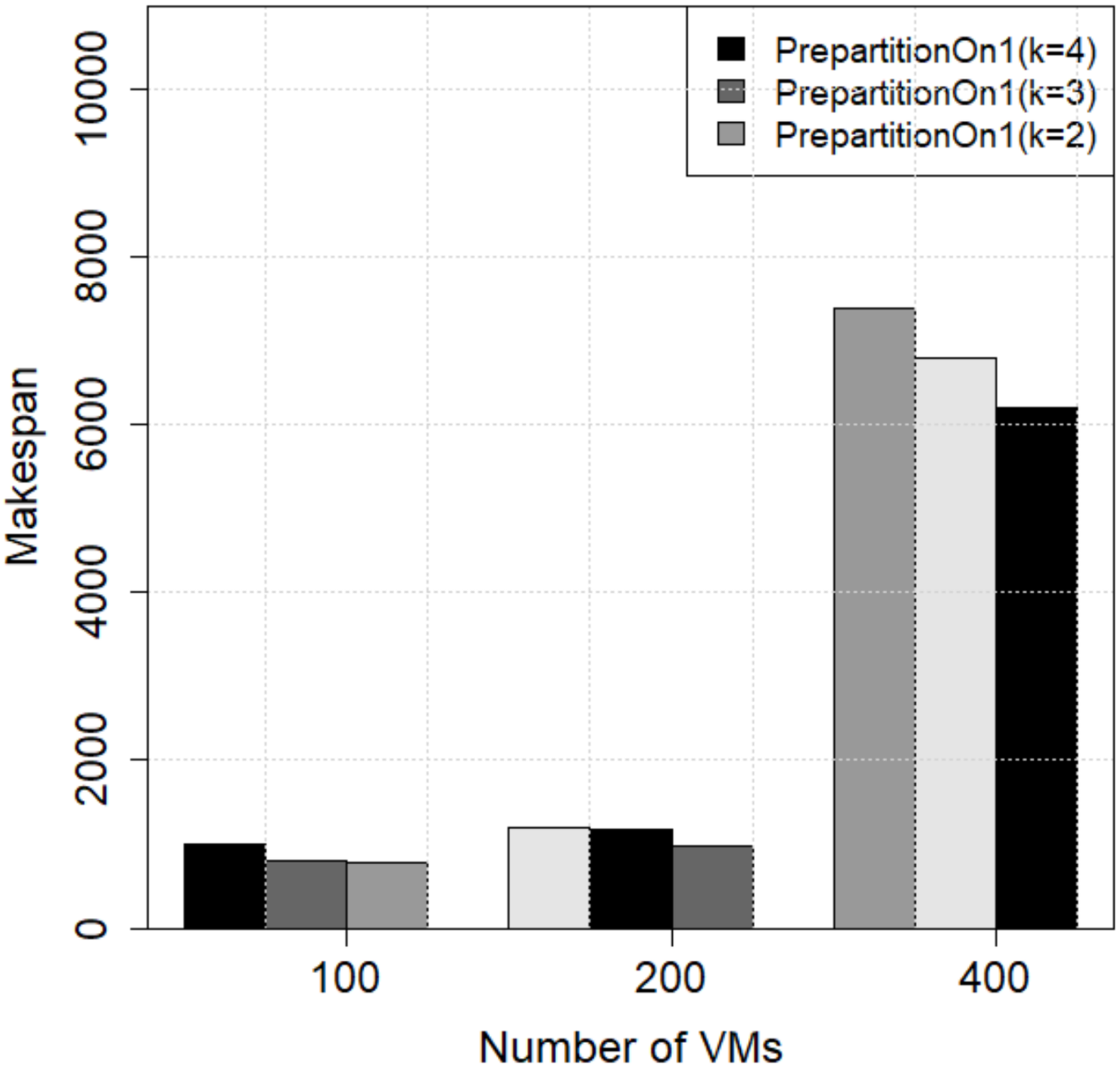}
	\parbox[c]{8.3cm}{\footnotesize{Fig.10.~}  The comparison of time costs for PrepartitionOn1 by varying $k$ values}
	\label{fig:kcomparison}
\end{center}
\setcounter{figure}{10}

\color{black}	To evaluate the number of partitions triggered by different Prepartition algorithms, Table \ref{tab:numerofmig} shows the number of partitions during our testes. Since the PrepartitionOff algorithm is offline, so the number is much smaller than the online algorithms. And the partitions of PrepartitionOn2 are smaller than PrepartitionOn1, as PrepartitionOn2 has brought predefined parameters to avoid too many partitions as discussed in Section 4.3.  \color{black}

\begin{table*}
	\color{black}
	\center
	\footnotesize
	\caption{\color{black}Number of partitions in different algorithms}
	\begin{tabular}{|c|c|c|}
		\hline
		\multirow{2}{*}{Algorithm} & \multicolumn{2}{|c|}{Number of partitions} \\
		\cline{2-3}
		& 1000 VMs & 2000 VMs \\
		\hline
		PrepartitionOff & 64 & 109\\
		\hline
		PrepartitionOn1 & 159 & 361\\
		\hline
		PrepartitionOn2 & 115 & 293 \\
		\hline
	\end{tabular}
	\label{tab:numerofmig}
\end{table*}

\section{Conclusions and Future Work}
\label{sec: Conclusion and Future Directions}
Load balancing for cloud administrators is a challenging problem in data centers. To address this issue, we proposed a novel virtual machine reservation paradigm to balance the VM loads for PMs. Through prepartition operations before allocation for VMs, our algorithm achieves better load balancing effects compared to well-known load balancing algorithms. In this paper, we present both offline and online load balancing algorithms to reveal the feature of fixed interval constraints of virtual machine scheduling and capacity sharing. Theoretically, we prove that PrepartitionOff is an algorithm with  (1+$\epsilon$) approximation ratio, where $\epsilon$=$\frac{1}{k}$ and $k$ is a positive integer. It is possible that the algorithm will be very close to the optimal solution via increasing the $k$ value, i.e., through setting up $k$, it is also attainable to achieve a desired load balancing goal defined in advance because PrepartitionOff is a (1+$\frac{1}{k}$)-approximation, PrepartitionOn1 has competitive ratio $(1+\frac{1}{k}-\frac{1}{mk})$ and PrepartitionOn2 has competitive ratio $(1+f)$ where $f$ is a constant below 0.5. Both the synthetic and trace-driven simulations have validated theoretical observations and shown that the Prepartition algorithms can perform better than a few existing algorithms at average utilization, imbalance degree, makespan, and Capacity$\_$makespan.
As such, other further research issues can be considered:
\begin{itemize}
	\item
	Making an appropriate choice between load balance and total partition numbers. Prepartition algorithm can achieve desired load balance objective by setting
	suitable $k$ value. It may need a large number of partitions so that the number of migrations can be large depending on the characteristics of VM requests. For
	example in EC2 \cite{IEEEhowto:Amazon}, the duration of VM reservations varies from a few hours to a few months, we can classify different types of VMs based on
	their durations (Capacity$\_$makespans) firstly, then applying Prepartition will not have a large partition number for each type.
	In practice, we need to analyze traffic patterns to make the number of partitions (pre-migrations) reasonable so that the total costs, including running time and
	the number migration can be reduced.
	\item
	Considering the heterogeneous configuration of PMs and VMs. We mainly consider that a VM requires a portion of the total capacity from a PM. This is also applied in EC2
	and Knauth et al. \cite{Knauth2012}. When this is not true, multi-dimensional resources, such as CPU, memory, and bandwidth, etc. have to be considered together or
	separately in the load balance, see \cite{IEEEhowto:Singh} and \cite{IEEEhowto:Sun} for a detailed discussion about considering multi-dimensional resources.
	\item
	\color{black}Considering precedence constraints among different VM requests. In reality, some VMs may be more important than others depending on applications running on them, we would like to extend current
	algorithm to consider this case.
	
	\item Considering the multi-tenancy and resource contention when making prepartitions, which can be investigated by characterizing application features. For instance, tightly coupled requests/applications can be partitioned on the same VM to reduce communication costs. 
	
	\color{black}
	
\end{itemize}
{\bf Acknowledgements}  The authors would like to thank the editors and anonymous reviewers' valuable comments to improve the quality of our work. This work is supported by Key-Area Research and Development Program of Guangdong Province (NO. 2020B010164003), National Natural Science Foundation of China (NO. 62102408) and SIAT Innovation Program for Excellent Young Researchers.





\bibliographystyle{unsrt}
\footnotesize	
\bibliography{jss-1}

\vspace{5mm}


\begin{wrapfigure}{l}{25mm} 
	\includegraphics[width=1in,height=1.25in,clip,keepaspectratio]{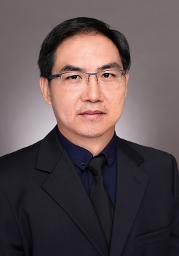}
\end{wrapfigure}\par
\textbf{Wenhong Tian} (Senior Member, CCF) has a PhD from Computer Science Department of North Carolina
State University. He is a full professor at University of Electronic Science and Technology
of China (UESTC). His research interests include resource scheduling in Cloud computing
and Bigdata processing, image identification and
classification, and automatic quality analysis of
natural language with deep learning. He published about 50 journal and conference papers,
and 3 English books in related areas. He is a
member of ACM, IEEE and CCF.  \par

	\vspace{10mm}
\begin{wrapfigure}{l}{25mm} 
	\includegraphics[width=1in,height=1.25in,clip,keepaspectratio]{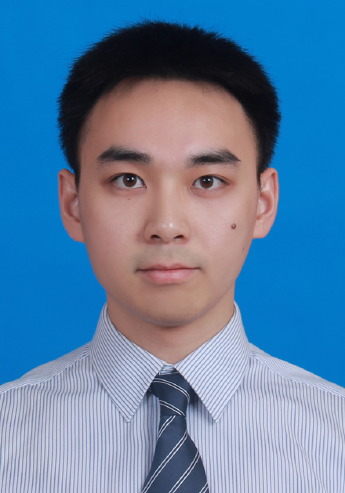}
\end{wrapfigure}\par
\textbf{Minxian Xu} (Member, IEEE) is currently an assistant professor at Shenzhen Institutes of Advanced Technology, Chinese Academy of Sciences. He received the BSc degree	in 2012 and the MSc degree in 2015, both in software engineering from University of Electronic Science and Technology of China. He obtained his PhD degree from the University of Melbourne in 2019. His research interests include resource scheduling and optimization in cloud computing. He has co-authored 28 peer-reviewed papers published in prominent international journals and conferences. His Ph.D. Thesis was awarded the 2019 IEEE TCSC Outstanding Ph.D. Dissertation Award.  \par

	\vspace{10mm}
	\begin{wrapfigure}{l}{25mm} 
		\includegraphics[width=1in,height=1.1in,clip,keepaspectratio]{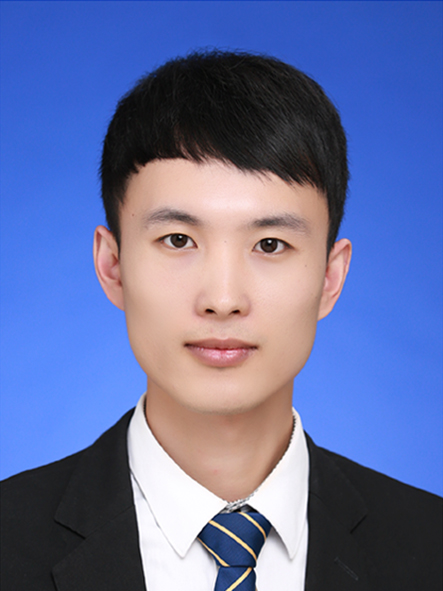}
	\end{wrapfigure}\par
	\textbf{Guangyao Zhou} received Bachelor's degree and Master's degree from School of architectural engineering, Tianjin University, China. He is now a Ph.D candidate at School of information and software engineering, University of Electronic Science and Technology of China, majoring in Software Engineering. His current research interests include Cloud Computing and image recognition.   \par

\vspace{10mm}
	\begin{wrapfigure}{l}{25mm} 
	\includegraphics[width=1in,height=1.25in,clip,keepaspectratio]{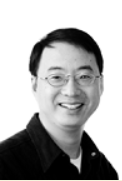}
\end{wrapfigure}\par
\textbf{Kui Wu} (Senior Member, IEEE) received the B.Sc. and M.Sc. degrees in computer science from Wuhan University, Wuhan, China, in 1990 and 1993, respectively, and the Ph.D. degree in computing science
from the University of Alberta, Edmonton, AB, Canada, in 2002. In 2002, he joined the Department of Computer Science, University of Victoria, Victoria, BC, Canada, where he is currently a professor. His current research interests include network performance analysis, online social networks, Internet of Things, and parallel and distributed algorithms.  \par

\vspace{5mm}
	\begin{wrapfigure}{l}{25mm} 
	\includegraphics[width=1in,height=1.25in,clip,keepaspectratio]{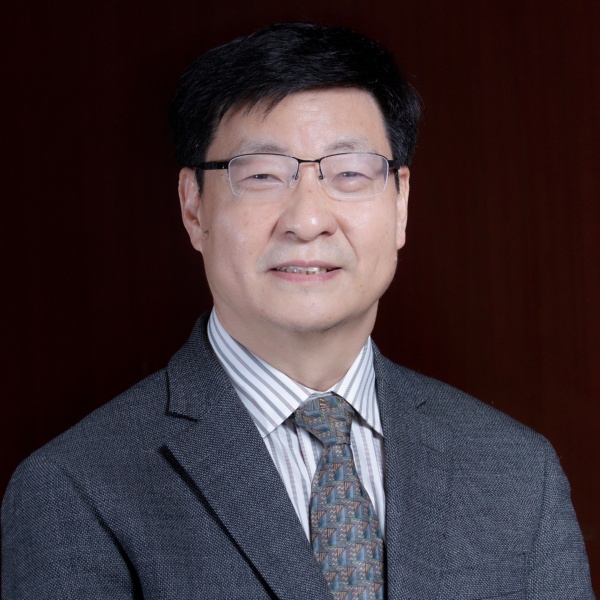}
\end{wrapfigure}\par
\textbf{Chengzhong Xu} (Fellow, IEEE) is the Dean of Faculty of Science and Technology and the Interim Director of Institute of Collaborative Innovation, University of Macau, and a Chair Professor of Computer and Information Science. Dr. Xu's main research interests lie in parallel and distributed computing and cloud computing, in particular, with an emphasis on resource management for system's performance, reliability, availability, power efficiency, and security, and in big data and data-driven intelligence applications in smart city and self-driving vehicles. He published two research monographs and more than 300 peer-reviewed papers in journals and conference proceedings; his papers received about 10K citations with an H-index of 52. 
He obtained BSc and MSc degrees from Nanjing University in 1986 and 1989 respectively, and a PhD degree from the University of Hong Kong in 1993, all in Computer Science and Engineering.  \par 
	
	\vspace{5mm}
	\begin{wrapfigure}{l}{25mm} 
		\includegraphics[width=1in,height=1.25in,clip,keepaspectratio]{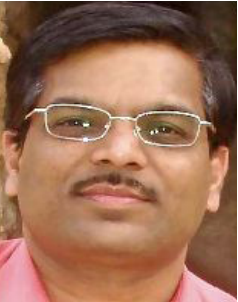}
	\end{wrapfigure}\par
	\textbf{Rajkumar Buyya} (Fellow, IEEE) is a Redmond Barry Distinguished Professor and Director of the Cloud Computing and Distributed Systems (CLOUDS) Laboratory at the University of Melbourne, Australia. 
	He has authored over 625 publications and seven text books.
	He is one of the highly cited authors in computer
	science and software engineering worldwide  (h-index=136, g-index=300, 98,800+ citations). Dr. Buyya is
	recognized as a "Web of Science Highly Cited Researcher" for four
	consecutive years since 2016, a Fellow of IEEE, and Scopus Researcher of the Year 2017 with Excellence in Innovative Research Award by Elsevier for his outstanding contributions to Cloud computing. 
	For further information on Dr. Buyya, please visit his cyberhome:
	www.buyya.com \par

\label{last-page}
\end{multicols}
\label{last-page}
\end{document}